\documentclass[runningheads]{llncs}
\usepackage[T1]{fontenc}
\usepackage{graphicx}
\usepackage[misc,geometry]{ifsym}
\usepackage{hyperref}
\usepackage{amsmath}
\usepackage{amssymb}
\usepackage{xcolor}
\usepackage{subcaption}
\definecolor{thetruth}{rgb}{0,0.701,0.117}

\definecolor{theboss}{rgb}{1.0,0.0,0.0}

\definecolor{arifaz}{rgb}{0.7, 0.3, 0.5}

\definecolor{new}{rgb}{0,0.7,1}

\usepackage{tikz}
\usetikzlibrary{arrows.meta}

\usepackage{xcolor}

\definecolor{lightergray}{rgb}{0.9, 0.9, 0.9}
\definecolor{darkgray}{rgb}{0.66, 0.66, 0.66}

\begin{document}

\title{Hypergraph p-Laplacians and Scale Spaces
\thanks{The authors acknowledge financial support by the European Unions Horizon 2020 research and innovation programme under the Marie Skodowska-Curie grant agreement No. 777826 (NoMADS) and the German Science Foundation (DFG) through CRC TR 154 ”Mathematical Modelling, Simulation and Optimization Using the Example of Gas Networks”, subproject C06. Part of this work was carried out while M. Burger was with the FAU Erlangen-Nürnberg.} }
\titlerunning{Hypergraph p-Laplacians and Scale Spaces}

\author{Ariane Fazeny\inst{1, 2}$^{\text{ \Letter}}$, Daniel Tenbrinck\inst{1}, Kseniia Lukin\inst{1}, Martin Burger\inst{2,3} }

\authorrunning{A.~Fazeny, D.~Tenbrinck, K.~Lukin, M.~Burger}

\institute{Friedrich-Alexander-Universität Erlangen-N\"urnberg, 91058 Erlangen, Germany \and Helmholtz Imaging, Deutsches Elektronen-Synchrotron DESY, \newline Notkestr. 85, 22607 Hamburg, Germany \and Universit\"at Hamburg, Fachbereich Mathematik, 20146 Hamburg, Germany\\ \Letter \,\email{ariane.fazeny@desy.de}}

\maketitle
\begin{abstract}
    The aim of this paper is to revisit the definition of differential operators on hypergraphs, which are a natural extension of graphs in systems based on interactions beyond pairs. {In particular, we focus on the definition of Laplacian and $p$-Laplace operators for oriented and unoriented hypergraphs,} their basic properties, variational structure, and their scale spaces.
    
    We illustrate that diffusion equations on hypergraphs are possible models for different applications such as information flow on social networks or image processing. Moreover, the spectral analysis and scale spaces induced by these operators provide a potential method to further analyze complex {data} and their multiscale structure. 

    The quest for spectral analysis and suitable scale spaces on hypergraphs motivates in particular a definition of differential operators with trivial first eigenfunction and thus more interpretable second eigenfunctions. This property is not automatically satisfied in existing definitions of hypergraph $p$-Laplacians and we hence provide a novel axiomatic approach that extends previous definitions and can be specialized to satisfy such (or other) desired properties.

    \keywords{Hypergraphs  \and PDEs on (hyper)graphs \and Diffusion models \and Information flow \and Hypergraph spectral clustering \and Image Processing \and Denoising \and Segmentation.}
\end{abstract}

\section{Introduction}

Methods for image processing, data analysis and simulation of information propagation have strongly benefited from using graph structures in the past, and the modeling with PDEs on graphs including graph $p$-Laplacians and associated flow became a standard tool for analyzing graph structures and dynamics on such (cf. \cite{elmoataz2015p,solomon2015,stankovic2019}). Those are carried on in machine learning in the concept of graph neural networks, again closely related to models for information flow on networks (cf. \cite{DiGiovanni}). Traditional graphs can however capture only  \textit{pairwise interactions} of individuals, objects, or pixels in images and thus are unable to directly model group relationships, which are relevant e.g. in social networks or image patches. In order to mitigate for this problem we propose to apply a more general structure, namely a \textbf{hypergraph} with which it is straightforward to encode group interactions. Here, we adopt the definition of oriented hypergraphs, whose hyperarcs (generalizing edges) can have more than one ingoing and more than one outgoing vertex. For this structure there is a natural way to define gradients, and we use a scaling which preserves the axiom that the gradient of a constant function on the vertices vanishes. Via a definition of adjoint we can then obtain a divergence operator and a Laplacian. Additionally, we investigate the case of unoriented hypergraphs, in which so-called hyperedges do not have an orientation, i.e., no destinction between outgoing and ingoing vertices. In contrast to traditional edges in graphs, here the number of vertices per hyperedge is not limited by two. For this type of hypergraph, we introduce two possible Laplacian operators, one which is gradient-based and one via an averaging operator.

\subsection{Motivation}

The hypergraph structure gives additional flexibility in several applications compared to the pair-based graph structure. An example is the modeling of social phenomena of (fake) news spread, e.g., by connecting one person to all their followers directly and hence representing a community within a social network. One field of study in analyzing information flow in social networks is \textit{opinion formation}, an interesting phenomenon that can be observed for a group of individuals which interact and have complex relationships with each other. 
Some individuals of the social network, so-called \textit{opinion leaders} or \textit{social media influencers}, with a large group of followers (up to half a billion people) have a strong influence on the opinion of many others and can even make profit by leveraging their impact on large groups of social media users (see, e.g., \cite{turcotte2015news}). 
Modeling information flow in social networks mathematically is typically performed by using traditional graphs. With such graphs it is possible to link two social media users with a pairwise connection, if they are online friends or follow each other (see, e.g., \cite{majeed2020graph}). The information flow in the social network can then be modeled in terms of diffusion processes on the graph, e.g., by solving a partial differential equation (PDE) {(see, e.g., \cite{arnaboldi2017online}, \cite{chamley2013models})}. However, recent work suggest that interactions beyond pairs are of particular relevance (cf. \cite{zanette}). Structures reminiscent of a Laplacian on hypergraphs can be found in the model of \cite{consensus}.

A similar question arises in the analysis of community structures, where graph spectral clustering is a standard technique. In order to understand networks including group connections, a more general structure such as hypergraphs seems to be more appropriate. The success of PDE-based methods on graphs motivates a further study on hypergraphs in order to explore the potential of PDEs on such objects. For this sake we need appropriate definitions of hypergraph gradients and Laplacians, which we revisit in this paper. Moreover, the study of scales on hypergraphs is a relevant topic, which could naturally be defined by the evolution of diffusion type processes we hence study here.

In image processing the graph structure is potentially limiting, since it merely confines to the comparison of pairs of pixels and their gray (respectively colour) values. It may however be relevant to make comparison between one pixel and its surrounding pixels without the restriction to all pairs. Another example is nonlocal image processing based on patches consisting of multiple images. Hypergraph p-Laplacians and their associated scale spaces are a promising approach for such.



\subsection{Related work}
There already exists extensive literature about traditional graph theory and its application to social networks. In \cite{majeed2020graph}, an overview of social network modeling with traditional graphs is given, including community clustering, similarity analysis and community-based event detection. It indicates how the versatile structure of a graph can be applied to real world problems. \cite{arnaboldi2017online} introduces the so-called ego network, a graph focusing on one specific social media user in the center and their surrounding concentric layers of followers, sorted hierarchically depending on their contact frequency. 

{The paper} \cite{elmoataz2015p} introduces first-order differential operators and a family of $p$-Laplacian operators for traditional oriented graphs.
The proposed partial difference, adjoint, divergence and anisotropic $p$-Laplacian for traditional graphs are a special case of our vertex gradient, adjoint, divergence and $p$-Laplacian operators for hypergraphs, which are introduced in Section \ref{diffOperators}.
The theoretical results of \cite{elmoataz2015p} are applied for mathematical image analysis, such as filtering, segmentation, clustering, and inpainting, but not for social network modeling.

\cite{jost2021plaplaceoperators} generalizes the already known $p$-Laplacian operators for normal graphs to the hypergraph setting and performs spectral analysis with a specific focus on the $1$-Laplacian. The spectral properties are then applied to common (hyper)graph problems, for instance vertex partitioning, cuts in graphs, coloring of vertices and hyperarc partitioning, but the paper does not include any numerical experiments or the modeling of social networks with hypergraphs.
In comparison, our gradient, adjoint, and $p$-Laplacian definitions are more general and also have the property of the gradient null space including constant vertex functions. Additionally, they are also more flexible with respect to their adaptability for application tasks.

\cite{consensus} uses unoriented hypergraphs to model different sociological phenomena of cliques, such as peer pressure, with consensus models. Diffusion processes in multi-way interactions with convergence to one united group consensus are modeled with a simple $2$-Laplacian inspired by the traditional graph setting. Due to the lack of orientation in the hypergraphs, the described consensus models are not able to capture the effects of a one-sided connection through following someone (e.g., Twitter, Instagram), but only mutual connection through being friends (e.g., Facebook).

Furthermore, \cite{zhou2006learning} uses unoriented hypergraphs in machine learning and shows how hypergraph modelling of data relationships can outperform normal graphs in  spectral clustering tasks. Similarly, \cite{li2018submodular} compares two different algorithms for submodular hypergraph clustering, for not oriented hypergraphs with positive vertex weights and a normalized positive hyperedge weight function, namely the Inverse Power Method (IPM) and the clique expansion method (CEM).

\subsection{Main contributions}
 
The contributions of this paper are manifold. First, we recall the generalized vertex $p$-Laplacian operators for oriented hypergraphs, which were introduced in the preceding paper \cite{fazeny2023} and \cite{masterarbeit}. They generalize the definitions in \cite{jost2021plaplaceoperators} by including two different vertex weight functions and hyperarc weight functions respectively. With appropriate choice of these weights the vertex gradient definition leading to the vertex $p$-Laplacian fulfills the expected properties of the continuum setting (antisymmetry and the gradient of a constant function being equal to zero), based on less strict assumptions compared to the implicit gradient of \cite{jost2021plaplaceoperators}. 

In order to obtain a meaningful definition of a $p$-Laplace operator on unoriented hypergraphs as well, we introduce a gradient operator with respect to a single vertex, which follows the idea of the respective operators in the oriented hypergraph case. As an alternative, we also consider an approach via an averaging operator on the unoriented hypergraph, which is however confined to the linear case ($p=2$) as of now. The two different Laplacian operators are subsequently compared in our numerical experiments.

Moreover, we include two possible applications of the corresponding diffusion equations: for the oriented setting we investigate the information flow on networks based on the hypergraph Laplacian and for the unoriented setting we discuss an application to image processing and derive novel scale spaces based on pixel neighbourhood comparison, for which our definitions are naturally suited.
\section{Mathematical basics of hypergraphs}

The definition of hypergraphs is a generalization of finite graphs, both in the case of unoriented and oriented hypergraphs, which are based on unoriented and oriented normal graphs respectively. Given a finite set of vertices $\mathcal{V} = \left\{v_1, v_2, \dots v_N\right\}$, then a hypergraph does not only capture pairwise connections between two vertices, but higher-order relationships within any subset of all vertices.

\begin{remark}
	As proposed in \cite{mulas2022random}, we differentiate between oriented and unoriented hypergraphs instead of directed and undirected hypergraphs, because for every oriented hyperarc there is only one orientation but two possible directions: the direction along the orientation and the direction against the orientation.
\end{remark}

\begin{definition}[\textbf{Unoriented hypergraph $UH$}]\label{UH} \cite{zhang2019signal}
	An unoriented hypergraph $UH = \left(\mathcal{V}, \mathcal{E}_H\right)$ consists of a finite set of vertices $\mathcal{V}$, and a set of so-called hyperedges $\mathcal{E}_H$, with each hyperarc $e_q \in \mathcal{E}_H$ being in the power set of the vertices $2^{\lvert \mathcal{V}\rvert}$ and satisfying $\emptyset \subset e_q \subset \mathcal{V}$ with $2 \leq \left\lvert e_q\right\rvert \leq \left\lvert \mathcal{V}\right\rvert - 1$.
\end{definition}

\begin{example}[\textbf{Unoriented hypergraph $UH$}]
    Given a set of vertices
	{\small \begin{equation*}
		\mathcal{V} = \left\{v_1, v_2, v_3, v_4, v_5, v_6, v_7, v_8\right\}
	\end{equation*}}
        \hspace{-0.6em} and a set of hyperedges
	{\small \begin{equation*}
		\mathcal{E}_H = \left\{\left\{v_1, v_2, v_5\right\}, \left\{v_2, v_3, v_7, v_8\right\}, \left\{v_6, v_7\right\}\right\},
	\end{equation*}}
        \hspace{-0.6em} then the unoriented hypergraph $UH = \left(\mathcal{V}, \mathcal{E}_H\right)$ can be visualized in the following way:
	
	\begin{center}\begin{tikzpicture}
			\tikzstyle{vertex} = [fill,shape=circle,node distance=40pt]
			\tikzstyle{edge} = [fill,opacity=.5,fill opacity=.5,line cap=round, line join=round, line width=30pt]
			\tikzstyle{elabel} =  [fill,shape=circle,node distance=40pt]
			
			\pgfdeclarelayer{background}
			\pgfsetlayers{background,main}
			
			\begin{scope}[every node/.style={circle,thick,draw}]
				\node (v1) at (0,0) {$v_1$};
				\node (v2) at (3,0) {$v_2$};
				\node (v3) at (6,0) {$v_3$};
				\node (v4) at (9, 0) {$v_4$};
				\node (v5) at (0,-3) {$v_5$};
				\node (v6) at (3,-3) {$v_6$};
				\node (v7) at (6,-3) {$v_7$};
				\node (v8) at (9,-3) {$v_8$};
			\end{scope}
			
			\begin{pgfonlayer}{background}
				\begin{scope}[transparency group,opacity=.2]
					\draw[edge,opacity=1,color=red] (v1.center) -- (v2.center) -- (v5.center) -- (v1.center);
					\fill[edge,opacity=1,color=red] (v1.center) -- (v2.center) -- (v5.center) -- (v1.center);
				\end{scope}
				\begin{scope}[transparency group,opacity=.2]
					\draw[edge,opacity=1,color=cyan] (v2.center) -- (v3.center) -- (v8.center) -- (v7.center) -- (v2.center);
					\fill[edge,opacity=1,color=cyan] (v2.center) -- (v3.center) -- (v8.center) -- (v7.center) -- (v2.center);
				\end{scope}
				\begin{scope}[transparency group,opacity=.2]
					\draw[edge,opacity=1,color=gray] (v6.center) -- (v7.center) -- (v6.center);
					\fill[edge,opacity=1,color=gray] (v6.center) -- (v7.center) -- (v6.center);
				\end{scope}
			\end{pgfonlayer}
			\node[text=red] (e1) at (1.1,-1.1) {$e_1$};
			\node[text=cyan] (e2) at (6,-1.5) {$e_2$};
			\node[text=gray] (e3) at (4.5,-3) {$e_3$};
	\end{tikzpicture}\end{center}\vspace{0.5cm}
	
	
			
			
			
\end{example}

\begin{remark}\label{uniqueE}
	For clarity reasons we assume that each hyperedge in  $\mathcal{E}_H$ is unique and hence occurs only once. This implies that the cardinality of the hyperedge set is finite due to set of vertices $\mathcal{V}$ being finite and the number of hyperedges in $UH = \left(\mathcal{V}, \mathcal{E}_H\right)$ being limited by $|\mathcal{E}_H| \leq N^N$.
\end{remark}

Assigning either an output or an input orientation to each vertex of a hyperedge results in an oriented version of hyperedges, so-called hyperarcs. Based on this, oriented hypergraphs can be defined.
\begin{definition}[\textbf{Oriented hypergraph $OH$}]\label{OH} \cite{jost2021plaplaceoperators}
	An oriented hypergraph $OH = \left(\mathcal{V}, \mathcal{A}_H\right)$ consists of a finite set of vertices $\mathcal{V}$, and a set of so-called hyperarcs $\mathcal{A}_H$. Each hyperarc $a_q \in \mathcal{A}_H$ contains two disjoint subsets of vertices
    {\small \begin{equation}
	a_q = \left(a_q^{out}, a_q^{in}\right)
    \end{equation}}
    \hspace{-0.8em} with $\emptyset \subset a_q^{out}, a_q^{in} \subset \mathcal{V}$, $a_q^{out} \cap a_q^{in} = \emptyset$, $a_q^{out}$ being the set of all output vertices and $a_q^{in}$ being the set of all input vertices of the hyperarc $a_q$.
\end{definition}
~\\[0.6cm]

\begin{example}[\textbf{Oriented hypergraph $OH$}]
    Given a set of vertices
	{\small \begin{equation*}
		\mathcal{V} = \left\{v_1, v_2, v_3, v_4, v_5, v_6, v_7, v_8\right\}
	\end{equation*}}
        \hspace{-0.6em} and a set of hyperarcs
	{\small \begin{equation*}
		\mathcal{A}_H = \left\{\left(\left\{v_1, v_2\right\}, \left\{v_5\right\}\right), \left(\left\{v_3, v_7\right\}, \left\{v_2, v_8\right\}\right), \left(\left\{v_6\right\}, \left\{v_7\right\}\right)\right\},
	\end{equation*}}
	\hspace{-0.6em} then the oriented hypergraph $OH = \left(\mathcal{V}, \mathcal{A}_H\right)$ can be visualized in the following way:
	
	\begin{center}\begin{tikzpicture}
			\tikzstyle{vertex} = [fill,shape=circle,node distance=40pt]
			\tikzstyle{edge} = [fill,opacity=.5,fill opacity=.5,line cap=round, line join=round, line width=30pt]
			\tikzstyle{elabel} =  [fill,shape=circle,node distance=40pt]
			
			\pgfdeclarelayer{background}
			\pgfsetlayers{background,main}
			
			\begin{scope}[every node/.style={circle,thick,draw}]
				\node (v1) at (0,0) {$v_1$};
				\node (v2) at (3,0) {$v_2$};
				\node (v3) at (6,0) {$v_3$};
				\node (v4) at (9, 0) {$v_4$};
				\node (v5) at (0,-3) {$v_5$};
				\node (v6) at (3,-3) {$v_6$};
				\node (v7) at (6,-3) {$v_7$};
				\node (v8) at (9,-3) {$v_8$};
			\end{scope}
			
			\begin{pgfonlayer}{background}
				\begin{scope}[transparency group,opacity=.2]
					\draw[edge,opacity=1,color=red] (v1.center) -- (v2.center) -- (v5.center) -- (v1.center);
					\fill[edge,opacity=1,color=red] (v1.center) -- (v2.center) -- (v5.center) -- (v1.center);
				\end{scope}
				\begin{scope}[transparency group,opacity=.2]
					\draw[edge,opacity=1,color=cyan] (v2.center) -- (v3.center) -- (v8.center) -- (v7.center) -- (v2.center);
					\fill[edge,opacity=1,color=cyan] (v2.center) -- (v3.center) -- (v8.center) -- (v7.center) -- (v2.center);
				\end{scope}
				\begin{scope}[transparency group,opacity=.2]
					\draw[edge,opacity=1,color=gray] (v6.center) -- (v7.center) -- (v6.center);
					\fill[edge,opacity=1,color=gray] (v6.center) -- (v7.center) -- (v6.center);
				\end{scope}
			\end{pgfonlayer}
			
			\node[text=red] (v1a1) at (0.5, -0.5) {$out$};
			\node[text=red] (v2a1) at (2.5, -0.5) {$out$};
			\node[text=red] (v5a1) at (0.5, -2.5) {$in$};
			\node[text=cyan] (v3a2) at (6, -0.75) {$out$};
			\node[text=cyan] (v2a2) at (3.5, -0.5) {$in$};
			\node[text=cyan] (v7a2) at (6, -2.25) {$out$};
			\node[text=cyan] (v8a2) at (8.5, -2.5) {$in$};
			\node[text=gray] (v6a3) at (3.75, -3) {$out$};
			\node[text=gray] (v7a3) at (5.25, -3) {$in$};
			\node[text=red] (a1) at (-1,-1.5) {$a_1$};
			\node[text=cyan] (a2) at (8.21,-0.79) {$a_2$};
			\node[text=gray] (a3) at (4.5,-4) {$a_3$};
	\end{tikzpicture}\end{center}
	
	Alternatively, hyperarcs can also be visualized similarly to arcs in normal graphs:
	
	\begin{center}\begin{tikzpicture}
			\tikzstyle{vertex} = [fill,shape=circle,node distance=80pt]
			\tikzstyle{edge} = [fill,opacity=.5,fill opacity=.5,line cap=round, line join=round, line width=40pt]
			\tikzstyle{elabel} =  [fill,shape=circle,node distance=40pt]
			
			\pgfdeclarelayer{background}
			\pgfsetlayers{background,main}
			
			\begin{scope}[every node/.style={circle,thick,draw}]
				\node (v1) at (0,0) {$v_1$};
				\node (v2) at (3,0) {$v_2$};
				\node (v3) at (6,0) {$v_3$};
				\node (v4) at (9, 0) {$v_4$};
				\node (v5) at (0,-3) {$v_5$};
				\node (v6) at (3,-3) {$v_6$};
				\node (v7) at (6,-3) {$v_7$};
				\node (v8) at (9,-3) {$v_8$};
			\end{scope}
			
			\begin{scope}[>={Stealth[red]},
				every node/.style={circle},
				every edge/.style={draw=red,very thick}]
				\path [->] (v1) edge node[] {} (v5);
				\path [->] (v2) edge node[] {} (v5);
			\end{scope}
			\begin{scope}[>={Stealth[cyan]},
				every node/.style={circle},
				every edge/.style={draw=cyan,very thick}]
				\path [->] (v3) edge node[] {} (v2);
				\path [->] (v3) edge node[] {} (v8);
                \path [->] (v7) edge node[] {} (v2);
				\path [->] (v7) edge node[] {} (v8);
			\end{scope}
			\begin{scope}[>={Stealth[gray]},
				every node/.style={circle},
				every edge/.style={draw=gray,very thick}]
				\path [->] (v6) edge node[] {} (v7);
			\end{scope}
			\node[text=red] (a1) at (0.75, -1.5) {$a_1$};
			\node[text=cyan] (a2) at (6, -1.5) {$a_2$};
			\node[text=gray, below] (a3) at (4.5,-3) {$a_3$};
	\end{tikzpicture}\end{center}

        {In our numerical experiments we will use the second visualization option (without color-coding the different hyperarcs) in order to simplify understanding of the links between vertices. Since the underlying oriented hypergraph will have a specific property ($ \left\lvert a_q^{out}\right\rvert = 1$), the examples will have a one-to-one correspondence between the ''normal graph visualization'' and the hypergraph visualization, which would generally not be given without color-coding each hyperarc.}
\end{example}

\begin{remark}
	Furthermore, for clarity reasons we assume that each hyperarc in the set of hyperarcs $\mathcal{A}_H$ is unique and thus occurs only once. This automatically implies that the cardinality of the hyperarc set is finite due to set of vertices $\mathcal{V}$ being finite. More precisely the number of hyperarcs in $OH = \left(\mathcal{V}, \mathcal{A}_H\right)$ is limited by $|\mathcal{A}_H| \leq N^N$.
\end{remark}

We now define different functions on both unoriented and oriented hypergraphs, which are used in Section \ref{diffOperators} to introduce differential operators inspired by the continuum setting. In order to efficiently denote whether vertex is part of a hyperedge for an unoriented hypergraph and to check if a vertex is part of a hyperarc as an output or an input vertex for an oriented hypergraph, we use different kinds of vertex-hyperedge and vertex-hyperarc characteristic functions.

\begin{definition}[\textbf{Vertex-hyperedge characteristic function $\delta$}]\label{vertexedgeindi}
	For an unoriented hypergraph $UH = \left(\mathcal{V}, \mathcal{E}_H\right)$, we define the vertex-hyperedge characteristic function $\delta$ as:
	{\small \begin{equation}
		\delta: ~ \mathcal{V} \times \mathcal{E}_H \longrightarrow \left\{0, 1\right\} \qquad 
		\left(v_i, e_q\right) \longmapsto \delta\left(v_i, e_q\right) = \left\{\begin{array}{ll}
			1 & \quad v_i \in e_q\\
			0 & \quad \text{otherwise}
		\end{array}\right..
	\end{equation}}
\end{definition}

\begin{definition}[\textbf{Vertex-hyperarc characteristic functions $\delta_{out}$, $\delta_{in}$}]\label{vertexarcindiOH}
	For an oriented hypergraph $OH = \left(\mathcal{V}, \mathcal{A}_H\right)$, we define the output vertex-hyperarc characteristic function $\delta_{out}$ as:
	{\small \begin{equation}
		\delta_{out}: ~ \mathcal{V} \times \mathcal{A}_H \longrightarrow \left\{0, 1\right\} \qquad 
		\left(v_i, a_q\right) \longmapsto \delta_{out}\left(v_i, a_q\right) = \left\{\begin{array}{ll}
			1 & \quad v_i \in a_q^{out}\\
			0 & \quad \text{otherwise}
		\end{array}\right..
	\end{equation}}
	\hspace{-0.8em} Respectively, the input vertex-hyperarc characteristic function $\delta_{in}$ is given by:
	{\small \begin{equation}
		\delta_{in}: ~ \mathcal{V} \times \mathcal{A}_H \longrightarrow \left\{0, 1\right\} \qquad
		\left(v_i, a_q\right) \longmapsto \delta_{in}\left(v_i, a_q\right) = \left\{\begin{array}{ll}
			1 & \quad v_i \in a_q^{in}\\
			0 & \quad \text{otherwise}
		\end{array}\right..
	\end{equation}}
    \hspace{-0.8em} Instead of defining separate $\delta_{out}$ and $\delta_{in}$ characteristic functions, it would also be possible to define one vertex-hyperarc characteristic function $\delta_*$ as:
    {\small \begin{equation}
		\delta_*: ~ \mathcal{V} \times \mathcal{A}_H \longrightarrow \left\{-1, 0, 1\right\} \qquad 
		\left(v_i, a_q\right) \longmapsto \delta_*\left(v_i, a_q\right) = \left\{\begin{array}{ll}
			-1 & \quad v_i \in a_q^{in}\\
                1 & \quad v_i \in a_q^{out}\\
			0 & \quad \text{otherwise}
		\end{array}\right..
	\end{equation}}
    \hspace{-0.8em} however, this would lead to more complex definitions of the vertex gradient, adjoint, and $p$-Laplacian operators later on, because it complicates weighing output and input vertices of a hyperarc differently.
\end{definition}

Real valued functions can be defined on the set of vertices $\mathcal{V}$, the set of hyperedges $\mathcal{E}_H$ and the set of hyperarcs $\mathcal{A}_H$ in order to link any kind of data to a hypergraph.

\begin{definition}[\textbf{Vertex functions $f$, and hyperedge or hyperarc functions $F$}]\label{funcionH}
    For both an unoriented hypergraph $UH = \left(\mathcal{V}, \mathcal{E}_H\right)$ and an oriented hypergraph $OH = \left(\mathcal{V}, \mathcal{A}_H\right)$, vertex functions are defined on the set of vertices as
	{\small \begin{equation}
		f: ~ \mathcal{V} \longrightarrow \mathbb{R} \qquad v_i \longmapsto f\left(v_i\right)
	\end{equation}}
    \hspace{-0.6em} with vertex weight functions being defined as
	{\small \begin{equation}
		w: ~ \mathcal{V} \longrightarrow \mathbb{R}_{> 0} \qquad v_i \longmapsto w\left(v_i\right).
	\end{equation}}
    \hspace{-0.65em} For an unoriented hypergraph $UH = \left(\mathcal{V}, \mathcal{E}_H\right)$, hyperedge functions are defined on the domain of the set of hyperedges as
	{\small \begin{equation}
		F: ~ \mathcal{E}_H \longrightarrow \mathbb{R} \qquad e_q \longmapsto F\left(e_q\right)
	\end{equation}}
	\hspace{-0.6em} with hyperedge weight functions being defined as
	{\small \begin{equation}
		W: ~ \mathcal{E}_H \longrightarrow \mathbb{R}_{> 0} \qquad e_q \longmapsto W\left(e_q\right).
	\end{equation}}
    \hspace{-0.8em} Similarly, for an oriented hypergraph $OH = \left(\mathcal{V}, \mathcal{A}_H\right)$, hyperarc functions are defined on the domain of the set of hyperarcs as
	{\small\begin{equation}
		F: ~ \mathcal{A}_H \longrightarrow \mathbb{R} \qquad a_q \longmapsto F\left(a_q\right)
	\end{equation}}
	\hspace{-0.66em} with hyperarc weight functions being defined as
	{\small\begin{equation}
		W: ~ \mathcal{A}_H \longrightarrow \mathbb{R}_{> 0} \qquad a_q \longmapsto W\left(a_q\right).
	\end{equation}}
\end{definition}



The space of all vertex functions, all hyperedge and all hyperarc functions defined on a given hypergraph can be identified with an $N$- or an at most $N^N$-dimensional Hilbert space, respectively.

\begin{definition}[\textbf{Space of vertex functions $\mathcal{H}\left(\mathcal{V}\right)$, space of hyperedge functions $\mathcal{H}\left(\mathcal{E}_H\right)$, and space of hyperarc functions $\mathcal{H}\left(\mathcal{A}_H\right)$}]\label{spacefuncH}
    For an unoriented hypergraph $UH = \left(\mathcal{V}, \mathcal{E}_H\right)$ and an oriented hypergraph $OH = \left(\mathcal{V}, \mathcal{A}_H\right)$, the space of all vertex functions $f$ is given by
	{\small \begin{equation}
		\mathcal{H}\left(\mathcal{V}\right) = \left\{f ~ \middle| ~ f: ~ \mathcal{V} \longrightarrow \mathbb{R}\right\}
	\end{equation}}
	\hspace{-0.8em} where $\mathcal{H}\left(\mathcal{V}\right)$ with the inner product ${\langle f, g \rangle}_{\mathcal{H}\left(\mathcal{V}\right)} = \sum_{v_i \in \mathcal{V}} w_I \left(v_i\right)^\alpha f\left(v_i\right) g\left(v_i\right)$ for any two vertex functions $f, g \in \mathcal{H}\left(\mathcal{V}\right)$, vertex weight function $w_I$, and parameter $\alpha \in \mathbb{R}$ is a Hilbert space.\\
    For an unoriented hypergraph $UH = \left(\mathcal{V}, \mathcal{E}_H\right)$, the space of all hyperedge functions $F$ is defined as
	{\small \begin{equation}
		\mathcal{H}\left(\mathcal{E}_H\right) = \left\{F ~ \middle| ~ F: ~ \mathcal{E}_H \longrightarrow \mathbb{R}\right\}
	\end{equation}}
     \hspace{-0.6em} where $\mathcal{H}\left(\mathcal{E}_H\right)$ with the inner product ${\langle F, G \rangle}_{\mathcal{H}\left(\mathcal{E}_H\right)} = \sum_{e_q \in \mathcal{E}_H} W_I \left(e_q\right)^\beta F\left(e_q\right) G\left(e_q\right)$ for any two hyperedge functions $F, G \in \mathcal{H}\left(\mathcal{E}_H\right)$, hyperedge weight function $W_I$, and parameter $\beta \in \mathbb{R}$ constitutes a Hilbert space.
    In the same manner, the space of all hyperarc functions $F$ for an oriented hypergraph $OH = \left(\mathcal{V}, \mathcal{A}_H\right)$ is defined as
    {\small \begin{equation}
		\mathcal{H}\left(\mathcal{A}_H\right) = \left\{F ~ \middle| ~ F: ~ \mathcal{A}_H \longrightarrow \mathbb{R}\right\}
	\end{equation}}
    \hspace{-0.8em} where $\mathcal{H}\left(\mathcal{A}_H\right)$ with the product ${\langle F, G \rangle}_{\mathcal{H}\left(\mathcal{A}_H\right)} = \sum_{a_q \in \mathcal{A}_H} W_I \left(a_q\right)^\beta F\left(a_q\right) G\left(a_q\right)$ for any two hyperarc functions $F, G \in \mathcal{H}\left(\mathcal{A}_H\right)$, hyperarc weight function $W_I$, and parameter $\beta \in \mathbb{R}$ defines a Hilbert space.
\end{definition}
\newpage
\section{Differential operators on hypergraphs}\label{diffOperators}

This section introduces first and higher order differential operators both for unoriented and oriented hypergraphs.

\subsection{First-order differential operators for oriented hypergraphs}\label{firstorderOH}

Utilizing the introduced definitions for hypergraphs we can now generalize the definitions of the vertex gradient, the vertex adjoint, and the vertex $p$-Laplacian for normal graphs, which have already been discussed in a simplified form with less weight functions and parameters in \cite{elmoataz2015p}.

\begin{definition}[\textbf{Vertex gradient operator $\nabla_v$}]\label{Hnabla_v}
	For an oriented hypergraph $OH = \left(\mathcal{V}, \mathcal{A}_H\right)$ with vertex weight functions $w_I$ and $w_G$, and hyperarc weight function $W_G$, we define the vertex gradient operator $\nabla_v$ with parameters $\alpha, \gamma, \epsilon, \eta \in \mathbb{R}$ as:
	{\small\begin{equation*}
		\nabla_v: ~ \mathcal{H}\left(\mathcal{V}\right) \longrightarrow \mathcal{H}\left(\mathcal{A}_H\right) \quad f \longmapsto \nabla_v f
	\end{equation*}
        \begin{equation*}
            \nabla_v f: ~ \mathcal{A}_H \longrightarrow \mathbb{R} \quad a_q \longmapsto \nabla_v f \left(a_q\right) = 
        \end{equation*}
	\begin{equation}
		W_G \left(a_q\right)^\gamma \sum_{v_i \in \mathcal{V}} \left(\delta_{in}\left(v_i, a_q\right) \frac{w_I \left(v_i\right)^\alpha w_G \left(v_i\right)^\epsilon}{\left\lvert a_q^{in}\right\rvert} - \delta_{out}\left(v_i, a_q\right) \frac{w_I \left(v_i\right)^\alpha w_G \left(v_i\right)^\eta}{\left\lvert a_q^{out}\right\rvert}\right) f\left(v_i\right).
	\end{equation}}

    The weight $w_I$ denotes the vertex weight function from the inner product of $\mathcal{H}\left(\mathcal{V}\right)$ and $w_G$ denotes the vertex weight function, which is introduced with the gradient operator. Using different values for the parameters $\epsilon$ and $\eta$ corresponds to putting different weights on the input and output vertices in the gradient of hyperarc $a_q$.
\end{definition}

The introduced vertex gradient fulfills two expected properties from the continuum setting, namely antisymmetry and the gradient of a constant function being equal to zero.

\begin{theorem}[\textbf{Vertex gradient operator properties}]\label{Hnabla_vprop}
	The vertex gradient $\nabla_v$ defined on an oriented hypergraphs $OH = \left(\mathcal{V}, \mathcal{A}_H\right)$ with vertex weight functions $w_I$ and $w_G$, and hyperarc weight function $W_G$, satisfies the following properties:
	\begin{itemize}
		\item[1)] {\bf Vanishing gradient of a constant vertex function:} If the condition
        {\small \begin{equation*}
            w_I \left(v_k\right)^\alpha w_G \left(v_k\right)^\epsilon = w_I \left(v_j\right)^\alpha w_G \left(v_j\right)^\eta
        \end{equation*}}
        \hspace{-0.7em} holds for all vertex combinations $v_j, v_k \in \mathcal{V}$ with $v_j \in a_q^{out}$ and $v_k \in a_q^{in}$ for a hyperarc $a_q \in \mathcal{A}_H$, then for every constant function $f$, i.e.
		$f\left(v_i\right) \equiv \overline{f}$ for all vertices $v_i \in \mathcal{V}$, we have $\nabla_v f \left(a_q\right) = 0$ for all hyperarcs $a_q \in \mathcal{A}_H$.
		\vspace{0.3cm}
		\item[2)] {\bf Antisymmetry:} Let $\epsilon = \eta$. Then the identity\\ $\nabla_v f \left(a_q^{out}, a_q^{in}\right) = - \nabla_v f \left(a_q^{in}, a_q^{out}\right)$ holds for all hyperarcs $a_q \in \mathcal{A}_H$. 
	\end{itemize}
\end{theorem}

\begin{proof}
    See \cite{masterarbeit} Theorem 9.2 (Vertex gradient operator properties).
\end{proof}

Let us mention one additional complication compared to the traditional graph case: While it is trivial to see that for a connected graph constant functions are the only elements in the nullspace of the gradient, this is not apparent for hypergraphs.

By computing the adjoint $\nabla^*_v$ of the vertex gradient we can introduce a consistent definition of a divergence operator on hypergraphs in analogy to traditional calculus. Detailed computation based on the relation
{\small\begin{equation}
		{\langle G, \nabla_v f \rangle}_{\mathcal{H}\left(\mathcal{A}_H\right)} = {\langle f, \nabla^*_v G \rangle}_{\mathcal{H}\left(\mathcal{V}\right)}
	\end{equation}}
	\hspace{-0.8em} for all vertex functions $f \in \mathcal{H}\left(\mathcal{V}\right)$ and all hyperarc functions $G \in \mathcal{H}\left(\mathcal{A}_H\right)$ can be found in \cite{masterarbeit} Theorem 9.9 (Connection vertex gradient $\nabla_v$ and vertex adjoint $\nabla_v^*$).

\begin{definition}[\textbf{Vertex adjoint operator $\nabla^*_v$}]\label{Hnabla^*_v} 
	For an oriented hypergraph $OH = \left(\mathcal{V}, \mathcal{A}_H\right)$ with vertex weight function $w_G$, and hyperarc weight functions $W_I$ and $W_G$, the vertex adjoint operator $\nabla^*_v$ with parameters $\beta, \gamma, \epsilon, \eta \in \mathbb{R}$ is given by:
	{\small\begin{equation*}
		\nabla^*_v: ~ \mathcal{H}\left(\mathcal{A}_H\right) \longrightarrow \mathcal{H}\left(\mathcal{V}\right) \quad F \longmapsto \nabla^*_v F
	\end{equation*}
        \begin{equation*}
            \nabla^*_v F: ~ \mathcal{V} \longrightarrow \mathbb{R} \quad v_i \longmapsto \nabla^*_v F \left(v_i\right) =
        \end{equation*}
	\begin{equation}
		\sum_{a_q \in \mathcal{A}_H} \left(\delta_{in}\left(v_i, a_q\right) \frac{w_G \left(v_i\right)^\epsilon}{\left\lvert a_q^{in}\right\rvert} - \delta_{out}\left(v_i, a_q\right) \frac{w_G \left(v_i\right)^\eta}{\left\lvert a_q^{out}\right\rvert}\right) W_I \left(a_q\right)^\beta W_G \left(a_q\right)^\gamma F\left(a_q\right).
	\end{equation}}
 
\end{definition}

\begin{definition}[\textbf{Vertex divergence operator $\mathrm{div}_v$}]\label{Hdiv_v}
	For an oriented hypergraph $OH = \left(\mathcal{V}, \mathcal{A}_H\right)$ with vertex weight function $w_G$, and hyperarc weight functions $W_I$ and $W_G$, the vertex divergence operator $\mathrm{div}_v$ with parameters $\beta, \gamma, \epsilon, \eta \in \mathbb{R}$ is given by:
	{\small\begin{equation*}
		\mathrm{div}_v: ~ \mathcal{H}\left(\mathcal{A}_H\right) \longrightarrow \mathcal{H}\left(\mathcal{V}\right) \quad F \longmapsto \mathrm{div}_v F
	\end{equation*}
        \begin{equation*}
            \mathrm{div}_v F: ~ \mathcal{V} \longrightarrow \mathbb{R} \quad v_i \longmapsto \mathrm{div}_v F \left(v_i\right) = - \nabla^*_v F \left(v_i\right) = 
        \end{equation*}
	\begin{equation}
		\sum_{a_q \in \mathcal{A}_H} \left(\delta_{out}\left(v_i, a_q\right) \frac{w_G \left(v_i\right)^\eta}{\left\lvert a_q^{out}\right\rvert} - \delta_{in}\left(v_i, a_q\right) \frac{w_G \left(v_i\right)^\epsilon}{\left\lvert a_q^{in}\right\rvert}\right) W_I \left(a_q\right)^\beta W_G \left(a_q\right)^\gamma F\left(a_q\right).
	\end{equation}}
\end{definition}

\subsection{$p$-Laplacian operators for oriented hypergraphs}\label{plapOH}

Based on the previous definitions we introduce a generalized vertex $p$-Laplacian inspired by the continuum setting, which implies that for all $p \in \left(1, \infty\right)$ and all vertex functions $f \in \mathcal{H}\left(\mathcal{V}\right)$ it holds true that:
{\small\begin{equation*}
	\Delta_v^p f = \mathrm{div}_v \left(\left\lvert \nabla_v f\right\rvert^{p - 2} \nabla_v f\right).
\end{equation*}}

Note that from the definition of the divergence as a negative adjoint of the gradient it becomes clear the oriented hypergraph $p$-Laplacian is the negative variation of the $p$-norm of the gradient, which allows to apply the full theory of eigenvalues of $p$-homogeneous functionals (see, \cite{bungertburger}). In particular, the oriented hypergraph Laplacian is a negative semidefinite linear operator and has a spectrum on the negative real line.

\begin{definition}[\textbf{Vertex $p$-Laplacian operator $\Delta_v^p$}]
	For an oriented hypergraph $OH = \left(\mathcal{V}, \mathcal{A}_H\right)$ with vertex weight functions $w_I$ and $w_G$, and with hyperarc weight functions $W_I$ and $W_G$, the vertex $p$-Laplacian operator $\Delta_v^p$ with parameters $\alpha, \beta, \gamma, \epsilon, \eta \in \mathbb{R}$ is given by:
	{\small\begin{equation*}
		\Delta_v^p: ~ \mathcal{H}\left(\mathcal{V}\right) \longrightarrow \mathcal{H}\left(\mathcal{V}\right) \quad f \longmapsto \Delta_v^p f \quad \quad \quad
		\Delta_v^p f: ~ \mathcal{V} \longrightarrow \mathbb{R} \quad v_i \longmapsto \Delta_v^p f \left(v_i\right) =
	\end{equation*}
	\begin{equation*}
		\sum_{a_q \in \mathcal{A}_H} \left(\delta_{out}\left(v_i, a_q\right) \frac{w_G \left(v_i\right)^\eta}{\left\lvert a_q^{out}\right\rvert} - \delta_{in}\left(v_i, a_q\right) \frac{w_G \left(v_i\right)^\epsilon}{\left\lvert a_q^{in}\right\rvert}\right) W_I \left(a_q\right)^\beta W_G \left(a_q\right)^{p \gamma} 
	\end{equation*}
	\begin{equation*}
		\left\lvert\sum_{v_j \in \mathcal{V}} \left(\delta_{in}\left(v_j, a_q\right) \frac{w_I \left(v_j\right)^\alpha w_G \left(v_j\right)^\epsilon}{\left\lvert a_q^{in}\right\rvert} - \delta_{out}\left(v_j, a_q\right) \frac{w_I \left(v_j\right)^\alpha w_G \left(v_j\right)^\eta}{\left\lvert a_q^{out}\right\rvert}\right) f\left(v_j\right)\right\rvert^{p - 2}
	\end{equation*}
	\begin{equation}
    \label{eq:p_laplace_oriented_hypergraph}
		\sum_{v_k \in \mathcal{V}} \left(\delta_{in}\left(v_k, a_q\right) \frac{w_I \left(v_k\right)^\alpha w_G \left(v_k\right)^\epsilon}{\left\lvert a_q^{in}\right\rvert} - \delta_{out}\left(v_k, a_q\right) \frac{w_I \left(v_k\right)^\alpha w_G \left(v_k\right)^\eta}{\left\lvert a_q^{out}\right\rvert}\right) f\left(v_k\right).
	\end{equation}}
\end{definition}

The following theorem states that the vertex $p$-Laplacian is well-defined.

\begin{theorem}[\textbf{Connection vertex gradient $\nabla_v$, vertex divergence $\mathrm{div}_v$, and vertex $p$-Laplacian $\Delta_v^p$}]
	For an oriented hypergraph $OH = \left(\mathcal{V}, \mathcal{A}_H\right)$ with vertex weight functions $w_I$ and $w_G$, and hyperarc weight functions $W_I$ and $W_G$, the vertex $p$-Laplacian $\Delta_v^p$ fulfills the equality
	{\small\begin{equation}
		\Delta_v^p f = \mathrm{div}_v \left(\left\lvert \nabla_v f\right\rvert^{p - 2} \nabla_v f\right)
	\end{equation}}
	\hspace{-0.6em} for all vertex functions $f \in \mathcal{H}\left(\mathcal{V}\right)$.
\end{theorem}

\begin{proof}
    See \cite{masterarbeit} Theorem 10.13 (Connection vertex divergence $\mathrm{div}_v$, vertex gradient $\nabla_v$, and vertex $p$-Laplacian $\Delta^p_v$).
\end{proof}

Moreover, our vertex $p$-Laplacian definition is a valid generalization of the definition introduced in \cite{jost2021plaplaceoperators}.

\begin{remark}[Parameter choice for the vertex $p$-Laplacian operator]\label{HDelta_v^pparam}
	The simplified definition of the vertex $p$-Laplacian introduced in \cite{jost2021plaplaceoperators} for any vertex function $f \in \mathcal{H}\left(\mathcal{V}\right)$ and for any vertex $v_i \in \mathcal{V}$ can be written in our notation as:
	{\small\begin{align}
		&\Delta_p f \left(v_i\right) = \frac{1}{\deg\left(v_i\right)} \sum_{\substack{a_q \in \mathcal{A}_H: ~ \delta_{out}\left(v_i, a_q\right) = 1 \\ \text{or} ~ \delta_{in}\left(v_i, a_q\right) = 1}} \left\lvert\sum_{v_j \in a_q^{in}} f\left(v_j\right) - \sum_{v_j \in a_q^{out}} f\left(v_j\right)\right\lvert^{p - 2} \nonumber \\
		& \qquad \qquad \left(\sum_{v_k \in \mathcal{V}} \left(\delta_{out}\left(v_i, a_q\right) \delta_{out}\left(v_k, a_q\right) + \delta_{in}\left(v_i, a_q\right) \delta_{in}\left(v_k, a_q\right)\right) f\left(v_k\right) - \right. \nonumber\\
		&  \qquad \qquad \left.\sum_{v_k \in \mathcal{V}} \left(\delta_{out}\left(v_i, a_q\right) \delta_{in}\left(v_k, a_q\right) + \delta_{in}\left(v_i, a_q\right) \delta_{out}\left(v_k, a_q\right)\right) f\left(v_k\right)\right) .
	\end{align}}
	\hspace{-0.8em} The factor $\left(\delta_{out}\left(v_i, a_q\right) \delta_{out}\left(v_k, a_q\right) + \delta_{in}\left(v_i, a_q\right) \delta_{in}\left(v_k, a_q\right)\right)$ is always equal to zero, unless $v_i, v_k \in a_q^{out}$ or $v_i, v_k \in a_q^{in}$, which means that the vertices $v_i$ and $v_k$ are co-oriented. Similarly, the factor $\left(\delta_{out}\left(v_i, a_q\right) \delta_{in}\left(v_k, a_q\right) + \delta_{in}\left(v_i, a_q\right) \delta_{out}\left(v_k, a_q\right)\right)$ ensures to only consider vertices $v_k \in \mathcal{V}$ which are anti-oriented compared to vertex $v_i$ and hence either $v_i \in a_q^{out}, v_k \in a_q^{in}$ or $v_i \in a_q^{in}, v_k \in a_q^{out}$.\\
	
	Thus, choosing the parameters of the vertex $p$-Laplacian $\Delta_v^p$ as $\alpha = 0$, $\beta = 0$, $\gamma = 0$, $\epsilon = 0$ and $\eta = 0$ together with excluding the $\frac{1}{\left\lvert a_q^{out}\right\rvert}$ and $\frac{1}{\left\lvert a_q^{in}\right\rvert}$ multiplicative factors and including a new $- \frac{1}{\deg\left(v_i\right)}$ factor in the vertex adjoint and the vertex divergence, results in the simplified vertex $p$-Laplacian introduced in \cite{jost2021plaplaceoperators}.\\
	
	Moreover, applying these parameter choices to the vertex gradient, the vertex adjoint and the vertex divergence leads to the following definitions for all vertex functions $f \in \mathcal{H}\left(\mathcal{V}\right)$, all hyperarc functions $F \in \mathcal{H}\left(\mathcal{A}_H\right)$, for all hyperarcs $a_q \in \mathcal{A}_H$ and all vertices $v_i \in \mathcal{V}$:
    {\small\begin{equation*}
        \nabla_v f \left(a_q\right) = \sum_{v_i \in \mathcal{V}} \left(\delta_{in}\left(v_i, a_q\right) - \delta_{out}\left(v_i, a_q\right)\right) f\left(v_i\right)
    \end{equation*}
    \begin{equation*}
        \nabla^*_v F \left(v_i\right) = - \frac{1}{\deg\left(v_i\right)} \sum_{a_q \in \mathcal{A}_H} \left(\delta_{in}\left(v_i, a_q\right) - \delta_{out}\left(v_i, a_q\right)\right) F\left(a_q\right)
    \end{equation*}
    \begin{equation*}
        \mathrm{div}_v \left(F\right) \left(v_i\right) = - \frac{1}{\deg\left(v_i\right)} \sum_{a_q \in \mathcal{A}_H} \left(\delta_{out}\left(v_i, a_q\right) - \delta_{in}\left(v_i, a_q\right)\right) F\left(a_q\right)
    \end{equation*}}
\end{remark}

\begin{proof}
    See \cite{masterarbeit} Theorem 10.12 (Parameter choice for the vertex $p$-Laplacian operator).
\end{proof}

Based on our axiomatic definition of the $p-$Laplacian via a gradient and adjoint divergence, it is straight-forward to verify its variational structure:

\begin{theorem}[$p$-Laplacian energy and derivatives]
    For an oriented hypergraph $OH = \left(\mathcal{V}, \mathcal{A}_H\right)$, the negative hypergraph $p$-Laplacian for $p \in \left(1, \infty\right)$ is the first variation of the associated $p$-Dirichlet energy
    {\small\begin{equation}
       E_p[f] := \sum_{v_i \in \mathcal{V}} |\nabla_v^p f \left(v_i\right)| ,
    \end{equation}}
    i.e. for every vertex function $f \in \mathcal{H}\left(\mathcal{V}\right)$ we have
     {\small\begin{equation}
       - \Delta_v^p f = E_p'[f].
    \end{equation}}
\end{theorem}
\subsection{First-order differential operators for unoriented hypergraphs}\label{diffOperatorsOH}

In order to retrieve a meaningful definition of a gradient also in the case of an unoriented hypergraph, where vertices in an hyperedge can not be separated into output and input vertices, we appoint for each hyperedge $e_q \in \mathcal{E}_H$ a specific vertex $v_{\tilde{q}} := v_i \in e_q$, which all other vertices in the hyperedge $v_j \in e_q \backslash\left\{v_{\tilde{q}}\right\}$ are compared to.

\begin{remark}
    If it is not clear how to choose a suitable special vertex $v_{\tilde{q}}$ for each hyperedge $e_q \in \mathcal{E}_H$ based on the application, then it is also possible to include each hyperedge $e_q$ exactly $\left\lvert e_q\right\rvert$ times in the set of hyperedges $\mathcal{E}_H$, where each version of the hyperedge has a different special vertex $v_i \in e_q$ (however, note that this requires paying particular attention to notation as mentioned in \ref{uniqueE}). This means that we are able to generate an oriented hypergraph out of an unoriented one as follows: For each hyperedge $e_q \in \mathcal{E}_H$ in the unoriented hypergraph, we create $|e_q|$ hyperarcs for the oriented hypergraph with the same vertices as $e_q$. Each hyperarc has one output and $|e_q|-1$ input vertices and each vertex $v_i \in e_q$ is an output vertex in exactly one newly created hyperarc. The associated vertex gradient, adjoint, and $p$-Laplacian operators for the oriented hypergraph then follow the respective definitions of the unoriented hypergraph case.
\end{remark}

Before defining new differential operators for unoriented hypergraphs, it is necessary to introduce a new vertex-hyperedge characteristic function.

\begin{definition}[\textbf{Vertex-hyperedge characteristic function $\tilde{\delta}$}]
	For an unoriented hypergraph $UH = \left(\mathcal{V}, \mathcal{E}_H\right)$, we define the vertex-hyperedge characteristic function $\tilde{\delta}$ as
	{\small \begin{equation}
		\tilde{\delta}: ~ \mathcal{V} \times \mathcal{E}_H \longrightarrow \left\{0, 1\right\} \qquad 
		\left(v_i, e_q\right) \longmapsto \tilde{\delta}\left(v_i, e_q\right) = \left\{\begin{array}{ll}
			1 & \quad v_i = v_{\tilde{q}}\\
			0 & \quad \text{otherwise}
		\end{array}\right.
	\end{equation}}
    \hspace{-0.8em} which indicates if vertex $v_i \in \mathcal{V}$ is the special vertex $v_{\tilde{q}}$ of hyperedge $e_q \in \mathcal{E}_H$. Furthermore, the following connection to the vertex-hyperedge function $\delta$ holds true for all vertices $v_i \in \mathcal{V}$ and all hyperedges $e_q \in \mathcal{E}_H$:
    {\small\begin{equation}
        \tilde{\delta}\left(v_i, e_q\right) = 1 \quad \Longrightarrow \quad \delta\left(v_i, e_q\right) = 1.
    \end{equation}}
\end{definition}

The vertex gradient operator for unoriented hypergraphs is defined with the same weight functions and parameters as the definition in the oriented case.

\begin{definition}[\textbf{Vertex gradient operator $\nabla_v$}]
	For an unoriented hypergraph $UH = \left(\mathcal{V}, \mathcal{E}_H\right)$ with vertex weight functions $w_I$ and $w_G$, and hyperedge weight function $W_G$, the vertex gradient operator $\nabla_v$ with parameters $\alpha, \gamma, \epsilon, \eta \in \mathbb{R}$ is given as:
	{\small\begin{equation*}
		\nabla_v: ~ \mathcal{H}\left(\mathcal{V}\right) \longrightarrow \mathcal{H}\left(\mathcal{E}_H\right) \quad f \longmapsto \nabla_v f
	\end{equation*}
        \begin{equation*}
            \nabla_v f: ~ \mathcal{E}_H \longrightarrow \mathbb{R} \quad e_q \longmapsto \nabla_v f \left(e_q\right) 
        \end{equation*}
        with 
        \begin{equation*}
        \begin{split}
            \nabla_v &f \left(e_q\right) = \\ &W_G \left(e_q\right)^\gamma \left(\sum_{v_i \in \mathcal{V}} \delta\left(v_i, e_q\right) \Big(w_I \left(v_i\right)^\alpha w_G \left(v_i\right)^\epsilon f\left(v_i\right) - w_I \left(v_{\tilde{q}}\right)^\alpha w_G \left(v_{\tilde{q}}\right)^\eta f\left(v_{\tilde{q}}\right)\Big)\right).
        \end{split}
        \end{equation*}
}
\end{definition}

The above gradient can be rewritten as:
	{\small\begin{equation}
    \begin{split}
		 \nabla_v &f \left(e_q\right) =\\ &W_G \left(e_q\right)^\gamma \left(\left(\sum_{v_i \in \mathcal{V}} \delta\left(v_i, e_q\right) w_I \left(v_i\right)^\alpha w_G \left(v_i\right)^\epsilon f\left(v_i\right)\right) - \left\lvert e_q\right\rvert w_I \left(v_{\tilde{q}}\right)^\alpha w_G \left(v_{\tilde{q}}\right)^\eta f\left(v_{\tilde{q}}\right)\right).
    \end{split}
	\end{equation}}
The vertex gradient for unoriented hypergraphs also fulfills the the expected property of a constant vertex function $f \in \mathcal{H}\left(\mathcal{V}\right)$ resulting in a vanishing gradient.

\begin{theorem}[\textbf{Vertex gradient operator properties}]
	The vertex gradient $\nabla_v$ of an unoriented hypergraph $UH = \left(\mathcal{V}, \mathcal{E}_H\right)$ with vertex weight functions $w_I$ and $w_G$, and hyperedge weight function $W_G$, satisfies the following property: If the vertex weights suffice the condition
    {\small\begin{equation*}
        w_I \left(v_i\right)^\alpha w_G \left(v_i\right)^\epsilon = w_I \left(v_{\tilde{q}}\right)^\alpha w_G \left(v_{\tilde{q}}\right)^\eta
    \end{equation*}}
    \hspace{-0.7em} for all vertex-hyperedge combinations $v_i \in e_q$ and $e_q \in \mathcal{E}_H$, then for every constant function $f \in \mathcal{H}\left(\mathcal{V}\right)$, i.e. $f\left(v_i\right) \equiv \overline{f}$ for all vertices $v_i \in \mathcal{V}$, we get $\nabla_v f \left(e_q\right) = 0$ for all hyperedges $e_q \in \mathcal{A}_H$.
\end{theorem}

\begin{proof}\ \\
    {Given a constant vertex function $f \in \mathcal{H}\left(\mathcal{V}\right)$ on an unoriented hypergraph $UH = \left(\mathcal{V}, \mathcal{E}_H\right)$ with vertex weight functions $w_I$ and $w_G$, and hyperedge weight function $W_G$, then it holds true that:
    The property $w_I \left(v_i\right)^\alpha w_G \left(v_i\right)^\epsilon = w_I \left(v_{\tilde{q}}\right)^\alpha w_G \left(v_{\tilde{q}}\right)^\eta$ for all vertex-hyperedge combinations $v_i \in e_q$ and $e_q \in \mathcal{E}_H$ implies that for each hyperedge $e_q$ there exists a constant $w_{e_q} \in \mathbb{R}_{> 0}$ such that
    {\small\begin{equation*}
	w_I \left(v_i\right)^\alpha w_G \left(v_i\right)^\epsilon = w_I \left(v_{\tilde{q}}\right)^\alpha w_G \left(v_{\tilde{q}}\right)^\eta =: w_{e_q}
    \end{equation*}}
    \hspace{-0.7em} for all vertices $v_i \in e_q$. Thus, together with the property $f\left(v_i\right) \equiv \overline{f} \in \mathbb{R}$ for all vertices $v_i \in \mathcal{V}$, this yields for every hyperedge $e_q \in \mathcal{E}_H$:\\}
    
    {\small $\begin{aligned}[t]	
	 \nabla_v f \left(e_q\right) 
        & = W_G \left(e_q\right)^\gamma \left(\sum_{v_i \in \mathcal{V}} \delta\left(v_i, e_q\right) \Big(w_I \left(v_i\right)^\alpha w_G \left(v_i\right)^\epsilon \overline{f} - w_I \left(v_{\tilde{q}}\right)^\alpha w_G \left(v_{\tilde{q}}\right)^\eta \overline{f}\Big)\right) &\\
        & = W_G \left(e_q\right)^\gamma \left(\sum_{v_i \in \mathcal{V}} \delta\left(v_i, e_q\right) \Big(w_I \left(v_i\right)^\alpha w_G \left(v_i\right)^\epsilon - w_I \left(v_{\tilde{q}}\right)^\alpha w_G \left(v_{\tilde{q}}\right)^\eta\Big) \overline{f}\right) &\\
        & = W_G \left(e_q\right)^\gamma \left(\sum_{v_i \in \mathcal{V}} \delta\left(v_i, e_q\right) \left(w_{e_q} - w_{e_q}\right) \overline{f}\right) &\\
        & = W_G \left(e_q\right)^\gamma \left\lvert e_q\right\rvert \left(w_{e_q} - w_{e_q}\right) \overline{f} &\\
        & = W_G \left(e_q\right)^\gamma \left\lvert e_q\right\rvert \cdot 0 \cdot \overline{f} = 0 &\\
	\end{aligned}$\\}
 
	Where the last equality is feasible due to the hyperedge weight function $W_G$ and the vertex function $\overline{f}$ being real functions and the number of vertices in every hyperedge $\left\lvert e_q\right\rvert$ being finite.\\
\end{proof}

Based on the connections in the continuum setting
\small{\begin{equation*}
    {\langle G, \nabla_v f \rangle}_{\mathcal{H}\left(\mathcal{E}_H\right)} = {\langle f, \nabla^*_v G \rangle}_{\mathcal{H}\left(\mathcal{V}\right)}
\end{equation*}
\begin{equation*}
    \mathrm{div}_v F = - \nabla^*_v F
\end{equation*}}
\hspace{-0.6em} for all vertex functions $f \in \mathcal{H}\left(\mathcal{V}\right)$ and all hyperedge functions $F, G \in \mathcal{H}\left(\mathcal{E}_H\right)$, we define the vertex adjoint and vertex gradient operators for unoriented hypergraphs.

\begin{definition}[\textbf{Vertex adjoint operator $\nabla^*_v$}]
	For an unoriented hypergraph $UH = \left(\mathcal{V}, \mathcal{E}_H\right)$ with vertex weight function $w_G$, and hyperedge weight functions $W_I$ and $W_G$, the vertex adjoint operator $\nabla^*_v$ with parameters $\beta, \gamma, \epsilon, \eta \in \mathbb{R}$ is given by:
	{\small\begin{equation*}
		\nabla^*_v: ~ \mathcal{H}\left(\mathcal{E}_H\right) \longrightarrow \mathcal{H}\left(\mathcal{V}\right) \quad F \longmapsto \nabla^*_v F
	\end{equation*}
        \begin{equation*}
            \nabla^*_v F: ~ \mathcal{V} \longrightarrow \mathbb{R} \quad v_i \longmapsto \nabla^*_v F \left(v_i\right) =
        \end{equation*}
	\begin{equation}
		\sum_{e_q \in \mathcal{E}_H} \left(\delta\left(v_i, e_q\right) w_G \left(v_i\right)^\epsilon - \tilde{\delta} \left(v_i, e_q\right) \left\lvert e_q\right\rvert w_G \left(v_i\right)^\eta \right) W_I \left(e_q\right)^\beta W_G \left(e_q\right)^\gamma F\left(e_q\right).
	\end{equation}}
\end{definition}

\begin{theorem}[\textbf{Connection vertex gradient $\nabla_v$ and vertex adjoint $\nabla^*_v$}]  \label{thm:adjoint}
    For an unoriented hypergraph $UH = \left(\mathcal{V}, \mathcal{E}_H\right)$ with vertex weight functions $w_I$ and $w_G$, and hyperedge weight functions $W_I$ and $W_G$, the vertex gradient $\nabla_v$ and the vertex adjoint $\nabla^*_v$ fulfill the equality
    {\small\begin{equation}
	{\langle G, \nabla_v f \rangle}_{\mathcal{H}\left(\mathcal{E}_H\right)} = {\langle f, \nabla^*_v G \rangle}_{\mathcal{H}\left(\mathcal{V}\right)}
    \end{equation}}
    \hspace{-0.6em} for all vertex functions $f \in \mathcal{H}\left(\mathcal{V}\right)$ and all hyperedge functions $G \in \mathcal{H}\left(\mathcal{E}_H\right)$.
\end{theorem}

\begin{proof}
    For the sake of clarity of this essay the proof is given in the appendix.
\end{proof}

As in the case of the oriented hypergraph, we define the vertex divergence operator based on the vertex adjoint operator.

\begin{definition}[\textbf{Vertex divergence operator $\mathrm{div}_v$}]
	For an unoriented hypergraph $UH = \left(\mathcal{V}, \mathcal{E}_H\right)$ with vertex weight function $w_G$, and hyperedge weight functions $W_I$ and $W_G$, the vertex divergence operator $\mathrm{div}_v$ with parameters $\beta, \gamma, \epsilon, \eta \in \mathbb{R}$ is given by:
	{\small\begin{equation*}
		\mathrm{div}_v: ~ \mathcal{H}\left(\mathcal{E}_H\right) \longrightarrow \mathcal{H}\left(\mathcal{V}\right) \quad F \longmapsto \mathrm{div}_v F
	\end{equation*}
        \begin{equation*}
            \mathrm{div}_v F: ~ \mathcal{V} \longrightarrow \mathbb{R} \quad v_i \longmapsto \mathrm{div}_v F \left(v_i\right) = - \nabla^*_v F \left(v_i\right) = 
        \end{equation*}
	\begin{equation}
		\sum_{e_q \in \mathcal{E}_H} \left(\tilde{\delta} \left(v_i, e_q\right) \left\lvert e_q\right\rvert w_G \left(v_i\right)^\eta - \delta\left(v_i, e_q\right) w_G \left(v_i\right)^\epsilon \right) W_I \left(e_q\right)^\beta W_G \left(e_q\right)^\gamma F\left(e_q\right).
	\end{equation}}
\end{definition}

\subsection{$p$-Laplacian operators for unoriented hypergraphs}\label{plapUH}

Analogously to the case of the oriented hypergraph, in this subsection we present a definition for the vertex $p$-Laplacian based on the vertex gradient and vertex divergence. The vertex Laplacian we obtain from a perspective of averaging, can be found in the next subsection.

\begin{definition}[\textbf{Vertex $p$-Laplacian operator $\Delta_v^p$}]
	For an unoriented hypergraph $UH = \left(\mathcal{V}, \mathcal{E}_H\right)$ with vertex weight functions $w_I$ and $w_G$, and hyperedge weight functions $W_I$ and $W_G$, the vertex $p$-Laplacian operator $\Delta_v^p$ with parameters $\alpha, \beta, \gamma, \epsilon, \eta \in \mathbb{R}$ is given by:
	{\small\begin{equation*}
		\Delta_v^p: ~ \mathcal{H}\left(\mathcal{V}\right) \longrightarrow \mathcal{H}\left(\mathcal{V}\right) \quad f \longmapsto \Delta_v f \quad \quad \quad
		\Delta_v^p f: ~ \mathcal{V} \longrightarrow \mathbb{R} \quad v_i \longmapsto \Delta_v^p f \left(v_i\right) =
	\end{equation*}}
    {\small\begin{equation*}
        \sum_{e_q \in \mathcal{E}_H} \left(\delta\left(v_i, e_q\right) w_G \left(v_i\right)^\epsilon - \tilde{\delta} \left(v_i, e_q\right) \left\lvert e_q\right\rvert w_G \left(v_i\right)^\eta \right) W_I \left(e_q\right)^\beta W_G \left(e_q\right)^{p \gamma}
    \end{equation*}}
    {\small\begin{equation*}
        \left\lvert \left(\sum_{v_j \in \mathcal{V}} \delta\left(v_j, e_q\right) w_I \left(v_j\right)^\alpha w_G \left(v_j\right)^\epsilon f\left(v_j\right)\right) - \left\lvert e_q\right\rvert w_I \left(v_{\tilde{q}}\right)^\alpha w_G \left(v_{\tilde{q}}\right)^\eta f\left(v_{\tilde{q}}\right)\right\rvert^{p - 2}
    \end{equation*}}
    {\small\begin{equation}\label{eq:p_laplace_unoriented_hypergraph}
        \left(\left(\sum_{v_k \in \mathcal{V}} \delta\left(v_k, e_q\right) w_I \left(v_k\right)^\alpha w_G \left(v_k\right)^\epsilon f\left(v_k\right)\right) - \left\lvert e_q\right\rvert w_I \left(v_{\tilde{q}}\right)^\alpha w_G \left(v_{\tilde{q}}\right)^\eta f\left(v_{\tilde{q}}\right)\right).
    \end{equation}}
\end{definition}

\begin{theorem}[\textbf{Connection vertex gradient $\nabla_v$, vertex divergence $\mathrm{div}_v$, and vertex $p$-Laplacian $\Delta_v^p$}]
	For an unoriented hypergraph $UH = \left(\mathcal{V}, \mathcal{E}_H\right)$ with vertex weight functions $w_I$ and $w_G$, and hyperedge weight functions $W_I$ and $W_G$, the presented vertex $p$-Laplacian $\Delta_v^p$ fulfills the equality
    {\small\begin{equation}
	\Delta_v^p f = \mathrm{div}_v \left(\left\lvert \nabla_v f\right\rvert^{p - 2} \nabla_v f\right)
    \end{equation}}
    \hspace{-0.6em} for all vertex functions $f \in \mathcal{H}\left(\mathcal{V}\right)$.
\end{theorem}

\begin{proof}\ \\
    {Given an unoriented hypergraph $UH = \left(\mathcal{V}, \mathcal{E}_H\right)$ with vertex weight functions $w_I$ and $w_G$, and hyperedge weight functions $W_I$ and $W_G$, and a vertex function $f \in \mathcal{H}\left(\mathcal{V}\right)$, then the definitions of the vertex divergence operator $\text{div}_v$ and the vertex gradient operator $\nabla_v$ lead to the following for all vertices $v_i \in \mathcal{V}$:}
	
    {\small $\begin{aligned}[t]	
	& \text{div}_v \left(\left\lvert \nabla_v f\right\rvert^{p - 2} \nabla_v f\right)\left(v_i\right) &\\
        = & \sum_{e_q \in \mathcal{E}_H} \left(\delta\left(v_i, e_q\right) w_G \left(v_i\right)^\epsilon - \tilde{\delta} \left(v_i, e_q\right) \left\lvert e_q\right\rvert w_G \left(v_i\right)^\eta \right) W_I \left(e_q\right)^\beta W_G \left(e_q\right)^\gamma &\\
        & \quad \left\lvert \nabla_v f \left(e_q\right)\right\rvert^{p - 2} \nabla_v f \left(e_q\right) &\\
        = & \sum_{e_q \in \mathcal{E}_H} \left(\delta\left(v_i, e_q\right) w_G \left(v_i\right)^\epsilon - \tilde{\delta} \left(v_i, e_q\right) \left\lvert e_q\right\rvert w_G \left(v_i\right)^\eta \right) W_I \left(e_q\right)^\beta W_G \left(e_q\right)^\gamma &\\
        & \quad \left\lvert W_G \left(e_q\right)^\gamma \left(\left(\sum_{v_j \in \mathcal{V}} \delta\left(v_j, e_q\right) w_I \left(v_j\right)^\alpha w_G \left(v_j\right)^\epsilon f\left(v_j\right)\right) - \left\lvert e_q\right\rvert w_I \left(v_{\tilde{q}}\right)^\alpha w_G \left(v_{\tilde{q}}\right)^\eta f\left(v_{\tilde{q}}\right)\right)\right\rvert^{p - 2} &\\
        & \quad W_G \left(e_q\right)^\gamma \left(\left(\sum_{v_k \in \mathcal{V}} \delta\left(v_k, e_q\right) w_I \left(v_k\right)^\alpha w_G \left(v_k\right)^\epsilon f\left(v_k\right)\right) - \left\lvert e_q\right\rvert w_I \left(v_{\tilde{q}}\right)^\alpha w_G \left(v_{\tilde{q}}\right)^\eta f\left(v_{\tilde{q}}\right)\right) &\\
    \end{aligned}$}\\

    Since the hyperedge weight function $W_G$ maps to positive values, the following equality holds true and leads to the vertex $p$-Laplacian definition for unoriented hypergraphs:
    
    {\small $\begin{aligned}[t]	
        = & \sum_{e_q \in \mathcal{E}_H} \left(\delta\left(v_i, e_q\right) w_G \left(v_i\right)^\epsilon - \tilde{\delta} \left(v_i, e_q\right) \left\lvert e_q\right\rvert w_G \left(v_i\right)^\eta \right) W_I \left(e_q\right)^\beta W_G \left(e_q\right)^{\gamma + \gamma \left(p - 2\right) + \gamma} &\\
        & \quad \left\lvert \left(\sum_{v_j \in \mathcal{V}} \delta\left(v_j, e_q\right) w_I \left(v_j\right)^\alpha w_G \left(v_j\right)^\epsilon f\left(v_j\right)\right) - \left\lvert e_q\right\rvert w_I \left(v_{\tilde{q}}\right)^\alpha w_G \left(v_{\tilde{q}}\right)^\eta f\left(v_{\tilde{q}}\right)\right\rvert^{p - 2} &\\
        & \left(\left(\sum_{v_k \in \mathcal{V}} \delta\left(v_k, e_q\right) w_I \left(v_k\right)^\alpha w_G \left(v_k\right)^\epsilon f\left(v_k\right)\right) - \left\lvert e_q\right\rvert w_I \left(v_{\tilde{q}}\right)^\alpha w_G \left(v_{\tilde{q}}\right)^\eta f\left(v_{\tilde{q}}\right)\right) &\\
        = & \sum_{e_q \in \mathcal{E}_H} \left(\delta\left(v_i, e_q\right) w_G \left(v_i\right)^\epsilon - \tilde{\delta} \left(v_i, e_q\right) \left\lvert e_q\right\rvert w_G \left(v_i\right)^\eta \right) W_I \left(e_q\right)^\beta W_G \left(e_q\right)^{p \gamma} &\\
        & \quad \left\lvert \left(\sum_{v_j \in \mathcal{V}} \delta\left(v_j, e_q\right) w_I \left(v_j\right)^\alpha w_G \left(v_j\right)^\epsilon f\left(v_j\right)\right) - \left\lvert e_q\right\rvert w_I \left(v_{\tilde{q}}\right)^\alpha w_G \left(v_{\tilde{q}}\right)^\eta f\left(v_{\tilde{q}}\right)\right\rvert^{p - 2} &\\
        & \left(\left(\sum_{v_k \in \mathcal{V}} \delta\left(v_k, e_q\right) w_I \left(v_k\right)^\alpha w_G \left(v_k\right)^\epsilon f\left(v_k\right)\right) - \left\lvert e_q\right\rvert w_I \left(v_{\tilde{q}}\right)^\alpha w_G \left(v_{\tilde{q}}\right)^\eta f\left(v_{\tilde{q}}\right)\right) &\\
        = & \Delta_v^p f \left(v_i\right)
    \end{aligned}$}\\
	
    Thus, the previously introduced definitions for the vertex gradient $\nabla_v$, the vertex divergence $\text{div}_v$, and the vertex $p$-Laplacian $\Delta_v^p$ suffice the equality $\Delta_v^p f\left(v_i\right) = \text{div}_v \left(\left\lvert \nabla_v f\right\rvert^{p - 2} \nabla_v f\right) \left(v_i\right)$ for all vertices $v_i \in \mathcal{V}$ and for all vertex functions $f \in \mathcal{H}\left(\mathcal{V}\right)$.\\
\end{proof}

\subsection{Averaging operators on unoriented hypergraphs}

Instead of starting with a gradient definition in order to retrieve a feasible Laplacian operator for unoriented hypergraphs, we now want to define a Laplacian operator based on intuitive averaging. For this definition a special vertex $v_{\tilde{q}}$ for every hyperedge $e_q \in \mathcal{E}_H$ is not necessary anymore. Before introducing the vertex averaging operator, we need the definition of the number of incident hyperedges.

\begin{definition}[\textbf{Number of incident hyperedges}]
    For an unoriented hypergraph $UH = \left(\mathcal{V}, \mathcal{E}_H\right)$, the number of incident hyperedges of a given vertex $v_i \in \mathcal{V}$ is defined as:
    {\small \begin{equation}
        \# \mathcal{E}_H \left(v_i\right) = \left\lvert\left\{e_q \in \mathcal{E}_H ~ \middle| ~ v_i \in e_q\right\}\right\rvert.
    \end{equation}}
    \hspace{-0.8em} Note: We call vertices $v_i \in \mathcal{V}$ with $\# \mathcal{E}_H \left(v_i\right) = 0$ isolated, since they are not connected to any other vertex $v_j \in \mathcal{V}$.
\end{definition}

The averaging operator below aims at defining for a given vertex $v_i$ the average value of a vertex function $f \in \mathcal{H}\left(\mathcal{V}\right)$ by considering all hyperedges $e_q \in \mathcal{E}_H$, which $v_i$ is a part of, and then averaging the vertex function over all vertices $v_j \in e_q$.

\begin{definition}[\textbf{Vertex averaging operator $\overline{\Delta_v}$}]
\label{def:averaging_operator}
	For an unoriented hypergraph $UH = \left(\mathcal{V}, \mathcal{E}_H\right)$ without any isolated vertices $v_i \in \mathcal{V}$, i.e. $\# \mathcal{E}_H \left(v_i\right) > 0$ for all vertices $v_i \in \mathcal{V}$, we define the vertex averaging operator as:
	{\small\begin{equation*}
		\overline{\Delta_v}: ~ \mathcal{H}\left(\mathcal{V}\right) \longrightarrow \mathcal{H}\left(\mathcal{V}\right) \quad f \longmapsto \overline{\Delta_v} f \quad \quad \quad
	\overline{\Delta_v} f: ~ \mathcal{V} \longrightarrow \mathbb{R} \quad v_i \longmapsto \overline{\Delta_v} f \left(v_i\right) =
	\end{equation*}
	\begin{equation}
		\frac{1}{\# \mathcal{E}_H\left(v_i\right)} \sum_{e_q \in \mathcal{E}_H} \delta\left(v_i, e_q\right) \frac{1}{\left\lvert e_q\right\rvert} \sum_{v_j \in \mathcal{V}} \delta\left(v_j, e_q\right) f \left(v_j\right).
	\end{equation}}
\end{definition}

By using a simplified version of the inner product on the space of all vertex functions $\mathcal{H}\left(\mathcal{V}\right)$ with $w_I \equiv 1$, we obtain an energy conserving adjoint vertex averaging operator $\overline{\Delta_v}^*$.

\begin{definition}[\textbf{Adjoint vertex averaging operator $\overline{\Delta_v}^*$}]
	For an unoriented hypergraph $UH = \left(\mathcal{V}, \mathcal{E}_H\right)$ without any isolated vertices and with the previously defined averaging operator $\overline{\Delta_v}$, the adjoint vertex averaging operator is given by:
	{\small\begin{equation*}
		\overline{\Delta_v}^*: ~ \mathcal{H}\left(\mathcal{V}\right) \longrightarrow \mathcal{H}\left(\mathcal{V}\right) \quad f \longmapsto \overline{\Delta_v}^* f \quad \quad \quad
		\overline{\Delta_v}^* f: ~ \mathcal{V} \longrightarrow \mathbb{R} \quad v_i \longmapsto \overline{\Delta_v}^* f \left(v_i\right) =
	\end{equation*}
	\begin{equation}
		\sum_{e_q \in \mathcal{E}_H} \delta\left(v_i, e_q\right) \frac{1}{\left\lvert e_q\right\rvert} \sum_{v_j \in \mathcal{V}} \delta\left(v_j, e_q\right) \frac{1}{\# \mathcal{E}_H\left(v_j\right)} f \left(v_j\right).
	\end{equation}}
\end{definition}

\begin{theorem}[\textbf{Connection between vertex averaging operator $\overline{\Delta_v}$ and adjoint vertex averaging operator $\overline{\Delta_v}^*$}]
	For an unoriented hypergraph $UH = \left(\mathcal{V}, \mathcal{E}_H\right)$ without isolated vertices and with any two vertex functions $f, g \in \mathcal{H}\left(\mathcal{V}\right)$, the vertex averaging operator $\Delta_v$ and the adjoint vertex operator $\Delta_v^*$ suffice the following equality
    {\small\begin{equation}
		{\langle g, \overline{\Delta_v} f \rangle}_{\mathcal{H}\left(\mathcal{V}\right)} := \sum_{v_i \in \mathcal{V}} g \left(v_i\right) \overline{\Delta_v} f \left(v_i\right) = \sum_{v_j \in \mathcal{V}} f \left(v_j\right) \overline{\Delta_v}^* g \left(v_j\right) =: {\langle f, \overline{\Delta_v}^* g \rangle}_{\mathcal{H}\left(\mathcal{V}\right)},
    \end{equation}}
    \hspace{-0.7em} where the inner product on the space of all vertex functions $\mathcal{H}\left(\mathcal{V}\right)$ has the weight $w_I \equiv 1$.
\end{theorem}

\begin{proof}\ \\
   Given an unoriented hypergraph $UH = \left(\mathcal{V}, \mathcal{E}_H\right)$ without any isolated vertices and two vertex functions $f, g \in \mathcal{H}\left(\mathcal{V}\right)$, then the definitions of the vertex averaging operator $\overline{\Delta_v}$ and the adjoint vertex averaging operator $\overline{\Delta_v}^*$ yield the following:\\
    
    {\small $\begin{aligned}[t]
	   {\langle g, \overline{\Delta_v} f \rangle}_{\mathcal{H}\left(\mathcal{V}\right)} = & \sum_{v_i \in \mathcal{V}} g\left(v_i\right) \overline{\Delta_v} f \left(v_i\right) &\\
        = & \sum_{v_i \in \mathcal{V}} g \left(v_i\right) \frac{1}{\# \mathcal{E}_H\left(v_i\right)} \sum_{e_q \in \mathcal{E}_H} \delta\left(v_i, e_q\right) \frac{1}{\left\lvert e_q\right\rvert} \sum_{v_j \in \mathcal{V}} \delta\left(v_j, e_q\right) f\left(v_j\right) &\\
        = & \sum_{v_i \in \mathcal{V}} \sum_{e_q \in \mathcal{E}_H} \sum_{v_j \in \mathcal{V}} g \left(v_i\right) \frac{1}{\# \mathcal{E}_H\left(v_i\right)} \delta\left(v_i, e_q\right) \frac{1}{\left\lvert e_q\right\rvert} \delta\left(v_j, e_q\right) f\left(v_j\right) &\\
        = & \sum_{v_j \in \mathcal{V}} \sum_{e_q \in \mathcal{E}_H} \sum_{v_i \in \mathcal{V}} g \left(v_i\right) \frac{1}{\# \mathcal{E}_H\left(v_i\right)} \delta\left(v_i, e_q\right) \frac{1}{\left\lvert e_q\right\rvert} \delta\left(v_j, e_q\right) f\left(v_j\right) &\\
        = & \sum_{v_j \in \mathcal{V}} f\left(v_j\right) \sum_{e_q \in \mathcal{E}_H} \delta\left(v_j, e_q\right) \frac{1}{\left\lvert e_q\right\rvert} \sum_{v_i \in \mathcal{V}} \delta\left(v_i, e_q\right) \frac{1}{\# \mathcal{E}_H\left(v_i\right)} g \left(v_i\right) &\\
        = & \sum_{v_j \in \mathcal{V}} f \left(v_j\right) \overline{\Delta_v}^* g \left(v_j\right) = {\langle f, 
        \overline{\Delta_v}^* g \rangle}_{\mathcal{H}\left(\mathcal{V}\right)} &\\
    \end{aligned}$\\}

    Therefore, with the definitions of the vertex averaging operator $\overline{\Delta_v}$ and the adjoint vertex averaging operator $\overline{\Delta_v}^*$, the equality ${\langle g, \overline{\Delta_v} f \rangle}_{\mathcal{H}\left(\mathcal{V}\right)} = {\langle f, \overline{\Delta_v}^* g \rangle}_{\mathcal{H}\left(\mathcal{V}\right)}$ holds true for all vertex functions $f, g \in \mathcal{H}\left(\mathcal{V}\right)$.\\
\end{proof}

\begin{example}[\textbf{Vertex averaging operator does not conserve mean values}]
   Given a set of vertices $\mathcal{V} = \left\{v_1, v_2, v_3, v_4\right\}$ and a set of hyperedges $\mathcal{E}_H = \left\{\left\{v_1, v_2, v_3\right\}, \left\{v_2, v_4\right\}\right\}$, then the vertex averaging operator $\overline{\Delta_v}$ on the unoriented hypergraph $UH = \left(\mathcal{V}, \mathcal{E}_H\right)$ does not conserve the mean value for general vertex functions $f \in \mathcal{H}\left(\mathcal{V}\right)$:
    
    \vspace{1cm}
    \begin{minipage}{.3\textwidth}\begin{tikzpicture}
	\tikzstyle{vertex} = [fill,shape=circle,node distance=40pt]
	\tikzstyle{edge} = [fill,opacity=.5,fill opacity=.5,line cap=round, line join=round, line width=30pt]
				
	\pgfdeclarelayer{background}
	\pgfsetlayers{background,main}
	\begin{scope}[every node/.style={circle,thick,draw}]
		\node (v1) at (0,2) {$v_1$};
		\node (v2) at (2,2) {$v_2$};
		\node (v3) at (0,0) {$v_3$};
		\node (v4) at (2,0) {$v_4$};
	\end{scope}
	\begin{pgfonlayer}{background}
		\begin{scope}[transparency group,opacity=.2]
			\draw[edge,opacity=1,color=red] (v1.center) -- (v2.center) -- (v3.center) -- (v1.center);
			\fill[edge,opacity=1,color=red] (v1.center) -- (v2.center) -- (v3.center) -- (v1.center);
		\end{scope}
		\begin{scope}[transparency group,opacity=.2]
			\draw[edge,opacity=1,color=cyan] (v2.center) -- (v4.center) -- (v2.center);
			\fill[edge,opacity=1,color=cyan] (v2.center) -- (v4.center) -- (v2.center);
		\end{scope}
	\end{pgfonlayer}
    \end{tikzpicture}\end{minipage}
    \begin{minipage}{.7\textwidth} 
    {\small $\overline{\Delta_v} f\left(v_1\right) = \frac{1}{1} \left(\frac{1}{3}\left(f\left(v_1\right) + f\left(v_2\right) + f\left(v_3\right)\right)\right)$\\
    
    $\overline{\Delta_v} f\left(v_2\right) = \frac{1}{2} \left(\frac{1}{3}\left(f\left(v_1\right) + f\left(v_2\right) + f\left(v_3\right)\right) + \frac{1}{2}\left(f\left(v_2\right) + f\left(v_4\right)\right)\right)$\\
    
    $\overline{\Delta_v} f\left(v_3\right) = \frac{1}{1} \left(\frac{1}{3}\left(f\left(v_1\right) + f\left(v_2\right) + f\left(v_3\right)\right)\right)$\\
    
    $\overline{\Delta_v} f\left(v_4\right) = \frac{1}{1} \left(\frac{1}{2}\left(f\left(v_2\right) + f\left(v_4\right)\right)\right)$}
    \end{minipage}

    \vspace{0.8cm}
    
    {\small $\overline{\Delta_v} f\left(v_1\right) + \overline{\Delta_v} f\left(v_2\right) + \overline{\Delta_v} f\left(v_3\right) + \overline{\Delta_v} f\left(v_4\right) = \frac{5}{6} f\left(v_1\right) + \frac{19}{12} f\left(v_2\right) + \frac{5}{6} f\left(v_3\right) + \frac{3}{4} f\left(v_4\right)$\\
    
    $\neq f\left(v_1\right) + f\left(v_2\right) + f\left(v_3\right) + f\left(v_4\right)$}\\
\end{example}

In contrast to this, the adjoint vertex averaging operator conserves the overall energy for all vertex functions $f \in \mathcal{H}\left(\mathcal{V}\right)$.

\begin{theorem}[\textbf{Adjoint vertex averaging operator $\overline{\Delta_v}^*$ conserves mean values}]
    For an unoriented hypergraph $UH = \left(\mathcal{V}, \mathcal{E}_H\right)$ without isolated vertices and with any vertex function $f \in \mathcal{H}\left(\mathcal{V}\right)$, the adjoint vertex averaging operator $\overline{\Delta_v}^*$ conserves the mean value of $f$, hence the following equality holds:
    {\small\begin{equation}
            \sum_{v_i \in \mathcal{V}} \overline{\Delta_v}^* f \left(v_i\right) = \sum_{v_i \in \mathcal{V}} f \left(v_i\right).
    \end{equation}}
\end{theorem}

\begin{proof}\ \\
    Given an unoriented hypergraph $UH = \left(\mathcal{V}, \mathcal{E}_H\right)$ without any isolated vertices and a vertex function $f \in \mathcal{H}\left(\mathcal{V}\right)$, then the following reformulations hold true:
	
	{\small $\begin{aligned}[t]	
		 \sum_{v_i \in \mathcal{V}} \overline{\Delta_v}^* f \left(v_i\right)
            = & \sum_{v_i \in \mathcal{V}} \sum_{e_q \in \mathcal{E}_H} \delta\left(v_i, e_q\right) \frac{1}{\left\lvert e_q\right\rvert} \sum_{v_j \in \mathcal{V}} \delta\left(v_j, e_q\right) \frac{1}{\# \mathcal{E}_H\left(v_j\right)} f \left(v_j\right) &\\
            = & \sum_{v_i \in \mathcal{V}} \sum_{e_q \in \mathcal{E}_H} \sum_{v_j \in \mathcal{V}} \delta\left(v_i, e_q\right) \frac{1}{\left\lvert e_q\right\rvert} \delta\left(v_j, e_q\right) \frac{1}{\# \mathcal{E}_H\left(v_j\right)} f \left(v_j\right) &\\
            = & \sum_{v_j \in \mathcal{V}} \sum_{e_q \in \mathcal{E}_H} \sum_{v_i \in \mathcal{V}} \delta\left(v_i, e_q\right) \frac{1}{\left\lvert e_q\right\rvert} \delta\left(v_j, e_q\right) \frac{1}{\# \mathcal{E}_H\left(v_j\right)} f \left(v_j\right) &\\
            = & \sum_{v_j \in \mathcal{V}} f \left(v_j\right) \frac{1}{\# \mathcal{E}_H\left(v_j\right)} \sum_{e_q \in \mathcal{E}_H} \delta\left(v_j, e_q\right) \frac{1}{\left\lvert e_q\right\rvert} \sum_{v_i \in \mathcal{V}} \delta\left(v_i, e_q\right) &\\
            = & \sum_{v_j \in \mathcal{V}} f \left(v_j\right) \frac{1}{\# \mathcal{E}_H\left(v_j\right)} \sum_{e_q \in \mathcal{E}_H} \delta\left(v_j, e_q\right) &\\
            = & \sum_{v_j \in \mathcal{V}} f \left(v_j\right) \frac{1}{\# \mathcal{E}_H\left(v_j\right)} ~ \# \mathcal{E}_H\left(v_j\right) &\\
            = & \sum_{v_j \in \mathcal{V}} f \left(v_j\right) &\\
	\end{aligned}$\\}
	
    Hence, the presented adjoint vertex averaging operator $\overline{\Delta_v}^*$ conserves the overall energy on any given unoriented hypergraph $UH$ for all vertex functions $f \in \mathcal{H}\left(\mathcal{V}\right)$.\\
\end{proof}

Let us put together the knowledge gained from the above analysis. Since we used the simple Euclidean scalar product and showed that the adjoint operator conserves the mean value, we can look at $ \overline{\Delta_v} - I$ as a suitable operator for a scale space analysis, somehow introducing a non-selfadjoint version of the Laplacian. The energy conservation of the adjoint shows that indeed $\overline{\Delta_v}$ has eigenvalue one with constant eigenfunction. Thus, the evolution equation with operator  $\overline{\Delta_v} - I$ is expected to converge to a constant state and yield a suitable scale space, which we will investigate further below.\\

Indeed, the averaging operator is self-adjoint if $\# \mathcal{E}_H\left(v_i\right)$ is constant on the set of vertices $\mathcal{V}$, and in this case $\overline{\Delta_v} - I$ has the structure of a normal graph Laplacian.

\begin{lemma}
Given an unoriented hypergraph $UH = \left(\mathcal{V}, \mathcal{E}_H\right)$ with $\# \mathcal{E}_H\left(v_i\right) = \# \mathcal{E}_H\left(v_j\right)$ for all vertices $v_i, v_j \in {\cal V}$. Then for all vertex functions $f \in \mathcal{H}\left(\mathcal{V}\right)$, the operator $\overline{\Delta_v} f - f$ is equivalent to the graph Laplacian on a weighted oriented graph for an arc $\left(v_i, v_j\right)$ from vertex $v_i$ to vertex $v_j$ with the particular weight function
$$ w(v_i,v_j) := \frac{1}{\# \mathcal{E}_H\left(v_i\right)} \sum_{e_q \in \mathcal{E}_H} \frac{1}{\left\lvert e_q\right\rvert} \delta\left(v_i, e_q\right) \delta\left(v_j, e_q\right).$$
\end{lemma}
\begin{proof}\ \\
For an unoriented hypergraph $UH = \left(\mathcal{V}, \mathcal{E}_H\right)$ with $\# \mathcal{E}_H\left(v_i\right) = \# \mathcal{E}_H\left(v_j\right)$ for all vertices $v_i, v_j \in {\cal V}$, the equivalence becomes apparent from a simple change of summation:
\begin{align*}
(\overline{\Delta_v} f -f) \left(v_i\right) =& \ \left(\frac{1}{\# \mathcal{E}_H\left(v_i\right)} \sum_{e_q \in \mathcal{E}_H} \delta\left(v_i, e_q\right) \frac{1}{\left\lvert e_q\right\rvert} \sum_{v_j \in \mathcal{V}} \delta\left(v_j, e_q\right) f(v_j) \right)- f(v_i) \\
=& \left(\frac{1}{\# \mathcal{E}_H\left(v_i\right)} \sum_{e_q \in \mathcal{E}_H} \delta\left(v_i, e_q\right) \frac{1}{\left\lvert e_q\right\rvert} \sum_{v_j \in \mathcal{V}} \delta\left(v_j, e_q\right) f(v_j)\right) &\\
-& \frac{1}{\# \mathcal{E}_H\left(v_i\right)} \# \mathcal{E}_H\left(v_i\right) \frac{1}{\left\lvert e_q\right\rvert} \left\lvert e_q\right\rvert f(v_i) \\
=& \left(\frac{1}{\# \mathcal{E}_H\left(v_i\right)} \sum_{e_q \in \mathcal{E}_H} \delta\left(v_i, e_q\right) \frac{1}{\left\lvert e_q\right\rvert} \sum_{v_j \in \mathcal{V}} \delta\left(v_j, e_q\right) f(v_j)\right) &\\
-& \left(\frac{1}{\# \mathcal{E}_H\left(v_i\right)} \sum_{e_q \in \mathcal{E}_H} \delta\left(v_i, e_q\right) \frac{1}{\left\lvert e_q\right\rvert} \sum_{v_j \in \mathcal{V}} \delta\left(v_j, e_q\right)\right) f(v_i)
\end{align*}
\begin{align*}
\hphantom{(\overline{\Delta_v} f -f) \left(v_i\right)} =& \frac{1}{\# \mathcal{E}_H\left(v_i\right)} \sum_{e_q \in \mathcal{E}_H} \delta\left(v_i, e_q\right) \frac{1}{\left\lvert e_q\right\rvert} \sum_{v_j \in \mathcal{V}} \delta\left(v_j, e_q\right) \left( f(v_j) - f(v_i)\right) \\
=& \sum_{e_q \in \mathcal{E}_H} \sum_{v_j \in \mathcal{V}} \frac{1}{\# \mathcal{E}_H\left(v_i\right)} \delta\left(v_i, e_q\right) \frac{1}{\left\lvert e_q\right\rvert} \delta\left(v_j, e_q\right) \left( f(v_j) - f(v_i)\right) \\
=& \sum_{v_j \in \mathcal{V}} \sum_{e_q \in \mathcal{E}_H} \left(\frac{1}{\# \mathcal{E}_H\left(v_i\right)} \delta\left(v_i, e_q\right) \frac{1}{\left\lvert e_q\right\rvert} \delta\left(v_j, e_q\right) \left( f(v_j) - f(v_i)\right)\right) \\
=& \sum_{v_j \in \mathcal{V}} \left(\frac{1}{\# \mathcal{E}_H\left(v_i\right)} \sum_{e_q \in \mathcal{E}_H} \frac{1}{\left\lvert e_q\right\rvert} \delta\left(v_i, e_q\right) \delta\left(v_j, e_q\right)\right) ( f (v_j) - f(v_i))\\
=& \sum_{v_j \in \mathcal{V}} w\left(v_i, v_j\right) ( f (v_j) - f(v_i))
\end{align*}

The term in the last row is exactly the traditional graph Laplace operator of vertex $v_i$ for an oriented normal graph with a vertex function $f$ (see, \cite{masterarbeit} Remark 7.12 (Parameter choice for the vertex $p$-Laplacian operator)), where for the arc weight $w$ it holds true that $w\left(v_i, v_j\right) = 0$ if the arc $\left(v_i, v_j\right)$ does not exist in the oriented normal graph.
\end{proof}

Let us mention that the weighted oriented normal graph we obtain above could be considered the easiest map from an unoriented hypergraph to a weighted graph, since the weights essentially count the number of hyperedges $e_q \in \mathcal{H}\left(\mathcal{V}\right)$ two vertices $v_i, v_j \in \mathcal{V}$ have in common.
\section{Scale spaces based on hypergraph $p$-Laplacians}\label{SSHY}

In the following we discuss PDEs based on the family of $p$-Laplace and averaging operators on hypergraphs introduced in Section \ref{diffOperators}, which can be used for  modeling information flow in social networks with oriented hypergraphs as well as performing image processing based on both oriented and unoriented hypergraphs.

\subsection{Modelling information flow using oriented hypergraphs}

For analyzing information flow on social networks with oriented hypergraphs, we consider two different PDE systems modelling diffusion processes. We start with investigating the scale space for the $p$-Laplacian operator, i.e. the gradient flow of the $p$-Laplacian energy:
\begin{equation}
\label{eq:pde_initial}
\begin{split}
\frac{\partial f}{\partial t}(v_i,t) \ &= \ \Delta_{v}^p f(v_i,t), \qquad v_i \in \mathcal{V}, t \in (0,\infty) \\
f(v_i,0) \ &= \ f_0(v_i), \hspace{1.26cm} v_i \in \mathcal{V} .
\end{split}
\end{equation}
Solving \eqref{eq:pde_initial} for every time step $t \in (0,\infty)$ amounts to computing the information flow between vertices of the oriented hypergraph along the respective hyperarcs. 


Note that although there are no explicit boundaries in oriented hypergraphs, we can interpret the above problem as the homogeneous Neumann boundary problem. Due to the properties of the proposed family of hypergraph $p$-Laplace operators it is easy to see that the mean-value of $f$ is conserved in time and we can naturally interpret the evolution as a scale space towards coarser and coarser scales on the graph. Moreover, the general asymptotic of gradient flows for $p$-homogeneous energies (cf. \cite{bungertburger}) yields that $f \rightarrow \overline{f}$ as $t \rightarrow \infty$, with $\overline{f}$ being the mean value of $f_0$. Moreover, the rescaled quantity
$    g = \frac{f - \overline{f}}{\Vert f - \overline{f} \Vert}$
converges to a multiple of a second eigenfunction for generic initial values.

Similar to the Neumann boundary problem, we can also introduce a Dirichlet type problem, where the Dirichlet boundary $\partial \mathcal{V} \subset \mathcal{V}$ denotes a subset of the vertex set $\mathcal{V}$ of the oriented hypergraph, for which we introduce boundary values and keep them fixed over time. The corresponding stationary solution is not necessarily constant 
\begin{equation}
\label{eq:pde_boundary}
\begin{split}
\Delta_{v}^p f(v_i) \ &= \ 0, \hspace{.88cm} v_i \in \mathring{\mathcal{V}},\\
f(v_j) \ &= \ F_j, \qquad v_j \in \partial\mathcal{V}.
\end{split}
\end{equation}
 
Then, we aim at solving the $p$-Laplace equation on the complementary vertex set $\mathring{\mathcal{V}} := \mathcal{V} \setminus \partial\mathcal{V}$ of the oriented hypergraph.
Instead of solving \eqref{eq:pde_boundary} directly, we solve the hyperbolic PDE model \eqref{eq:pde_initial} on the vertex set $\mathring{\mathcal{V}}$, while keeping the vertex function $f \in \mathcal{H}(\mathcal{V})$ fixed on the boundary set $\partial \mathcal{V}$.
The reason for this approach is that any stationary solution of \eqref{eq:pde_initial} on $\mathring{\mathcal{V}}$ with fixed boundary values is also a solution to the $p$-Laplace equation in \eqref{eq:pde_boundary}.
To solve the two proposed PDE models discussed above, we numerically have to solve the initial value problem in \eqref{eq:pde_initial}.
For this sake we employ a forward-Euler time discretization with fixed time step size $\tau > 0$ and use the renormalized variable $g$ to observe convergence to a nontrivial eigenfunction.
This leads to the following explicit iteration scheme:
\begin{equation}
\label{eq:iteration_scheme}
f_{n+1}(v_i) \ = \ f_{n}(v_i) + \tau \cdot \Delta_v^p f_n(v_i).
\end{equation}



\subsection{Image processing using unoriented hypergraphs}
To perform image processing for grayscale images defined on regular grids, we consider a PDE system that can be interpreted as an initial value problem for the vertex averaging operator introduced in Definition \ref{def:averaging_operator}. In particular, we are interested in solving the following initial value problem

\begin{equation}
\label{eq:initial_value_problem_averaging}
\begin{split}
\overline{\Delta_{v}} f \left(v_i, t\right) - f\left(v_i, t\right) + \lambda \cdot (f_0\left(v_i\right) - f\left(v_i, t\right)) \ &= \ 0, \qquad \qquad v_i \in \mathcal{V}, t \in (0,\infty) \\
f\left(v_i, 0\right) \ \ &= \ f_0\left(v_i\right), \hspace{0.55cm} v_i \in \mathcal{V}.
\end{split}
\end{equation}

Note that in contrast to the initial value problem modeling opinion formation in social networks in \eqref{eq:pde_initial}, here we introduce an additional data fidelity term that penalizes strong deviations from the noisy image, represented by the initial vertex function $f_0 \in \mathcal{H}(\mathcal{V})$.
The influence of this data fidelity term can be controlled by a fixed parameter $\lambda > 0$ that allows to realize a trade-off between smoothing the perturbed image pixels via the hypergraph vertex averaging operator and staying close to the initial image. In classical variational regularization this would correspond to the gradient flow of the least-squares fidelity augmented with a regularization energy scaled with regularization parameter $\frac{1}\lambda$. However, since the averaging operator is not self-adjoint the corresponding term in  (\ref{eq:initial_value_problem_averaging}) cannot arise in the gradient flow of an associated energy functional. Nonetheless, the diffusive nature of $\overline{\Delta_{v}} f   - f$ induces an interpretation as regularization albeit in nonvariational setting, similar e.g. to the inpainting model in \cite{cahnhilliard}. 

Once again we use a forward-Euler time discretization with fixed time step size $\tau > 0$, which is chosen small enough to fulfill the CFL stability conditions. Following this approach we derive the following iterative scheme for image processing using the hypergraph vertex averaging operator:

\begin{equation}
\label{eq:iterative_scheme_averaging}
f_{n+1}\left(v_i\right) = f_n\left(v_i\right) + \tau \cdot \left(\overline{\Delta_{v}}f_n(v_i) - f_n\left(v_i\right) + \lambda \cdot(f_0\left(v_i\right) - f_n\left(v_i\right))\right) .
\end{equation}
\section{Numerical experiments}
\label{s:numerics}
In this section we present the results of our numerical experiments when using the hypergraph operators introduced in Section \ref{diffOperators} for two different applications. In particular, we first discuss how the oriented hypergraph $p$-Laplacian operator can be used to model opinion formation in social networks. Furthermore, we apply the vertex averaging operator of unoriented hypergraphs for the task of image processing, and we provide results that can be used, both for image denoising as well as segmentation tasks.

\subsection{Opinion formation in social networks}
In the following we present the results of our numerical experiments in which we solve the two PDEs \eqref{eq:pde_initial} and \eqref{eq:pde_boundary} by using the explicit forward-Euler discretization scheme until the relative change between two iterations is smaller than $\epsilon := 10^{-6}$.
We choose $\tau$ in \eqref{eq:iteration_scheme} small enough to fulfill the CFL condition for numerical stability. This leads to very small time steps for the iteration scheme in the cases $1 \leq p < 2$.


For our numerical experiments we use the Twitter data set provided by Stanford University \cite{snapnets}. It consists of $41,652,230$ vertices (users) and $1,468,364,884$ arcs (oriented pairwise connections indicating that one person follows another).
Due to the size of the data set, we restrict our numerical experiments to a comparatively small sub-network within the first $1,000,000$ lines of the Twitter input data. We chose a sub-network of individuals such that all users are directly or indirectly linked to each other to avoid cliques of individuals, which are not connected to the rest of the sub-network and thus also not influenced by users outside their small circle.
Moreover, we ensure that each sub-network includes an opinion leader with a large number of followers in the sub-network. Therefore, we can observe how one influential user impacts the opinion of the rest.
In order to generate hyperarcs from the given arcs, we put one Twitter user as a singleton output vertex set and summarize all followers of this user as the set of input vertices. This especially allows highlighting the effect of opinion leaders, for instance famous people with a large group of followers on Twitter.

We simulate the opinion of all individuals in the social network towards an imaginary hypothesis by a vertex function $f \colon \mathcal{V} \times [0, \infty) \rightarrow [-1,1]$, which can be interpreted as the following.
If an individual believes the hypothesis the corresponding value of the vertex function is positive (with $1$ being the strongest level of trust), while for an individual that opposes the hypothesis the corresponding value of the vertex function is negative (with $-1$ being the strongest level of distrust).

\begin{figure}[t]
\includegraphics[width=\textwidth]{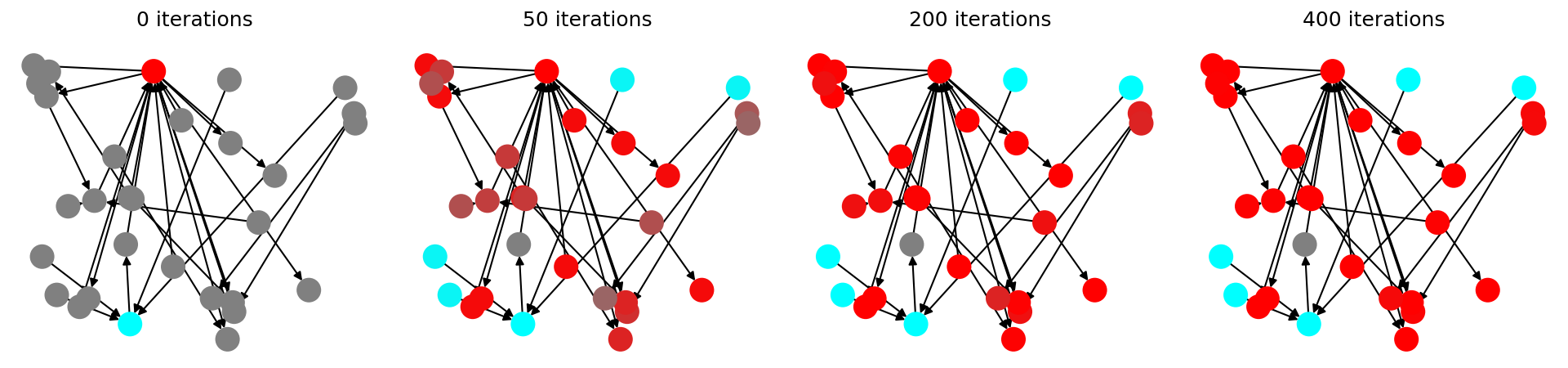}\\
\includegraphics[width=\textwidth]{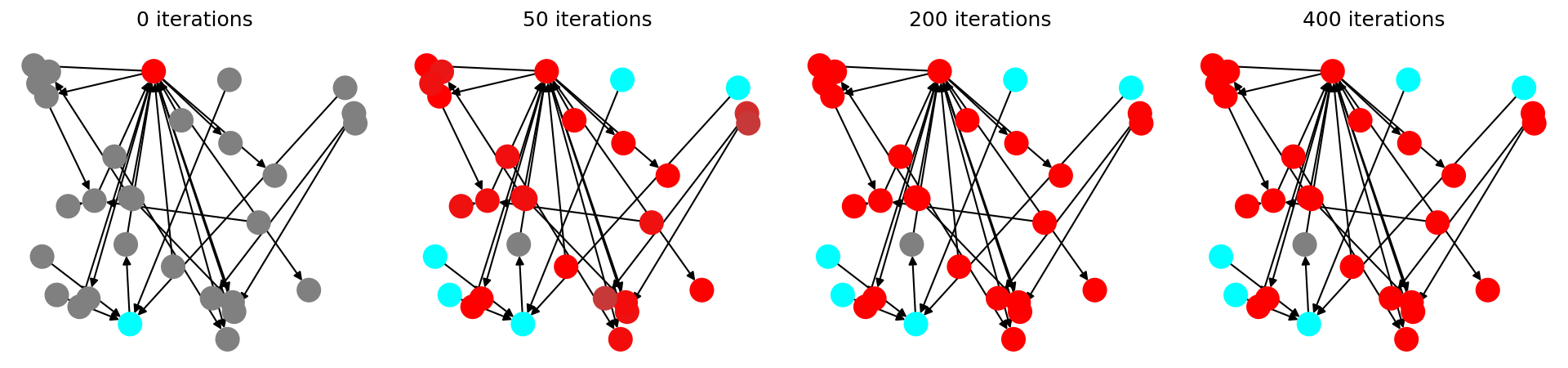}
\caption{Solution of the boundary value problem of graph (top) and hypergraph (bottom) $p$-Laplace operator for $p=2$.}
\label{fig:information_flow}
\end{figure}
For the \textbf{boundary value problem} \eqref{eq:pde_boundary} we initialize the opinion of all individuals in a social network by setting the vertex function $f$ to zero, which can be interpreted as having no opinion towards an imaginary hypothesis.
We now simulate information flow in the social network by giving two opinion leaders (i.e., vertices with many followers) two opposing opinions towards this hypothesis and setting the respective values of the vertex function to $-1$ and $1$.
We keep these values fixed as a form of Dirichlet boundary conditions.
By using the explicit forward-Euler discretization scheme to solve the boundary value problem for $p=2$, the opinion of the two dedicated individuals is propagated in the social network as can be seen in Figure \ref{fig:information_flow}.
We initialize the vertex function equally for the oriented normal graph (top row) and the oriented hypergraph (bottom row) and calculate the diffusion process until convergence. 
As can be seen, in both cases the opinion is propagated in the social network based on the underlying network topology and the final state is equivalent for both the normal graph and the hypergraph experiment.
However, as can be observed, information within the hypergraph is distributed at a higher rate compared to the normal graph and thus converging faster.
This is due to the fact that opinion leaders in a normal graph have a less direct impact on their followers compared to the hypergraph case, where the follower's believe $f(v_i)$ is scaled with $\frac{1}{\lvert a_q^{in}\rvert}$, where $\lvert a_q^{in}\rvert$ is the number of followers of the individual user.
This can be seen in \eqref{eq:p_laplace_oriented_hypergraph} since in our modeling for this application the parameter $a_q^{out}$ is set to $1$.

\begin{figure}[t]
\includegraphics[width=\textwidth]{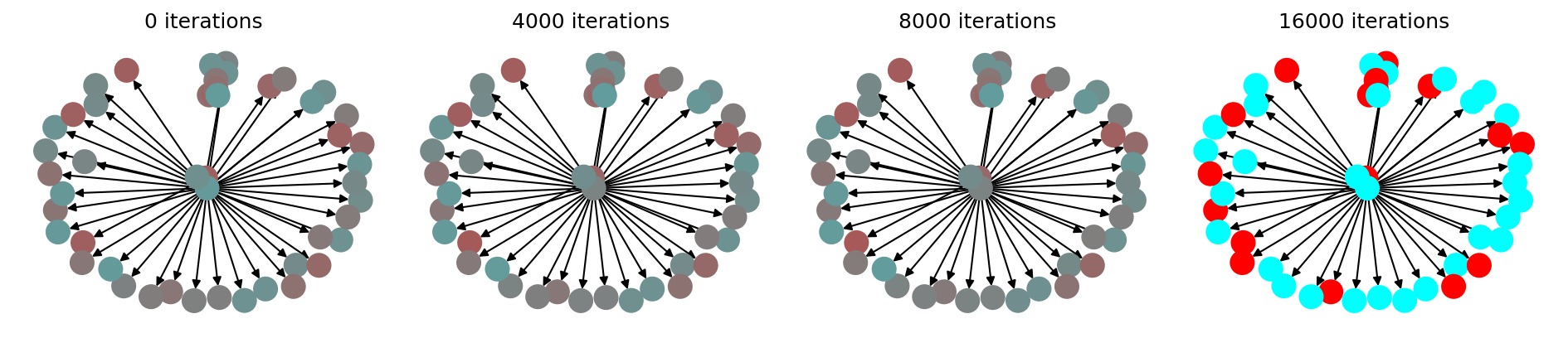}\\
\includegraphics[width=\textwidth]{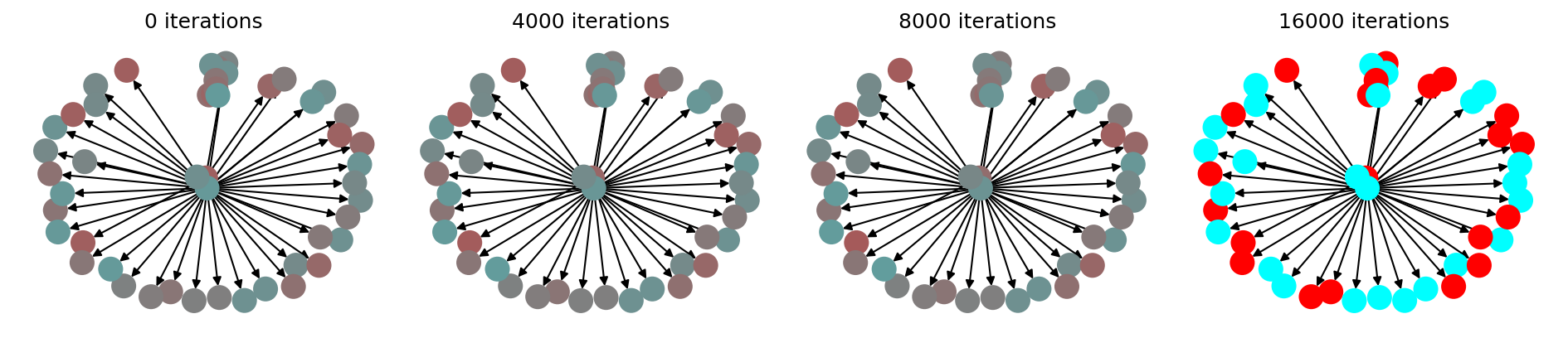}
\caption{Second eigenfunction of graph (top) and hypergraph (bottom) $p$-Laplace operator for $p=1$ with thresholding at $0$ after $16.000$ iterations.}
\label{fig:eigenfunction}
\end{figure}

For the \textbf{initial value problem} \eqref{eq:pde_initial} we choose $p=1$ {and a sufficiently small time step size $\tau > 0$ to guarantee stability of the corresponding iteration scheme \eqref{eq:iteration_scheme}}. 
We initialize each individual's opinion $f_0\left(v_i\right)$ randomly with a uniform distribution in the interval $\left[-1, 1\right]$.
Additionally, we make sure that the vertex function $f$ is initialized with average $0$ and normalized.
As can be observed in Figure \ref{fig:eigenfunction}, the information flow in the social network converges to a second eigenfunction of both the graph $p$-Laplacian (top row) and the hypergraph $p$-Laplacian (bottom row).
For both cases we thresholded at $0$ after $16.000$ iterations to induce a spectral clustering of the opposing opinions and hence separating the social network into smaller communities based on the topology of the network (i.e., the relationship of following an individual).
The resulting second eigenfunctions differ significantly with respect to the underlying topology of the oriented normal graph and the oriented hypergraph.
This yields potential for further analysis and experiments in other applications, e.g., segmentation of images via spectral clustering.


\subsection{Local and nonlocal image processing}
In the following, we discuss how the proposed hypergraph differential operators can be applied to image processing tasks.
By modeling pixels of an image with the help of normal graphs or hypergraphs instead of a regular grid, it is possible to not only represent local relationships of adjacent pixels, but also nonlocal relationships based on the image's content. For example, one could link image pixels that are relatively far from each other in the image, but share a similar image texture in their respective neighborhood.

Given an image $\tilde{I} \in \mathbb{R}^{n \times m}$ of height $n\in \mathbb{N}$ and width $m \in \mathbb{N}$  perturbed by a normal distributed noise signal $\nu \sim \mathcal{N}(0, \sigma^2)$, a typical task in image processing is to recover a noise-free image $I \in \mathbb{R}^{n \times m}$ from
\begin{equation}
\tilde{I} \ = \ I + \nu.
\end{equation}
This task can be interpreted as an inverse problem known as \textit{image denoising}.
In our numerical experiments we restrict ourselves to grayscale images for the sake of simplicity.

To perform image denoising we first model the relationship between the image pixels with an unoriented hypergraph.
In particular, we represent each image pixel as a vertex $v_i \in \mathcal{V}$ of the oriented hypergraph and interpret the pixel grayscale intensities as the values of the vertex function $f \in \mathcal{H}(\mathcal{V})$ with $f \colon \mathcal{V} \rightarrow [0,255]$. Here, $0$ represents the lowest signal intensity (i.e., black pixels) and $255$ represents the highest signal intensity (i.e., white pixels).
We construct the hyperedges of the hypergraph as described in detail below.

\begin{figure}
\includegraphics[width=0.48\textwidth]{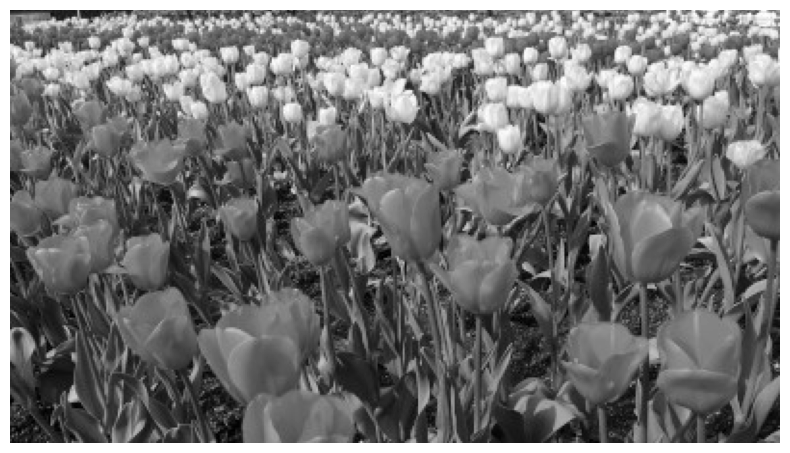}\hfill
\includegraphics[width=0.48\textwidth]{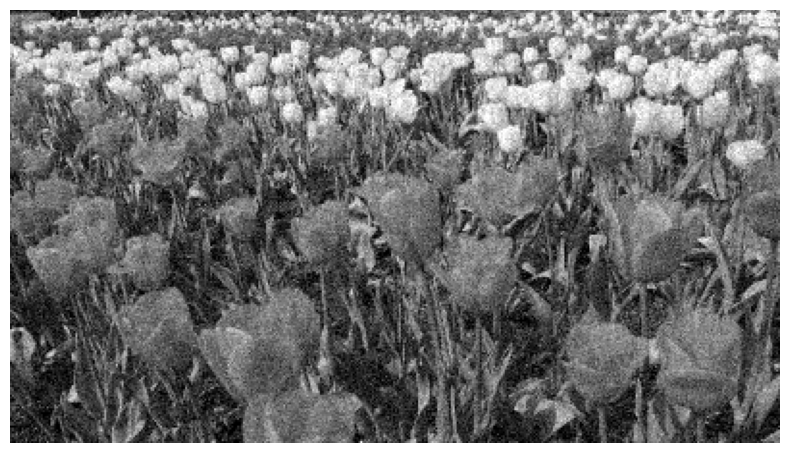}
\caption{Illustration of the original grayscale image $I$ (left) and the artificially perturbed image $\tilde{I}$ (right) used for image denoising.}
\label{fig:images}
\end{figure}

For our numerical experiments we chose a grayscale image $I$ of size $225 \times 400$ pixels of a flower field that contains image features at different scales.
We added random Gaussian noise $\nu$ with mean $\mu = 0$ and variance $\sigma^2 = 150$ to every image pixel to generate an artificially perturbed image $\tilde{I}$.
Both, the unperturbed image $I$ as well as the noisy variant $\tilde{I}$, are illustrated in Figure \ref{fig:images}
We construct an initial vertex function $f_0 \in \mathcal{H}(\mathcal{V})$ from the noisy image $\tilde{I}$.

We performed two different experiments for image denoising using the introduced unoriented hypergraph vertex average operator $\overline{\Delta_v}$ from definition \ref{def:averaging_operator}.
In the first experiment we perform \textbf{local image processing} by constructing a hyperedge of the unoriented hypergraph from the vertices that model the direct four pixel neighbors of any image pixel.
This corresponds to traditional image processing methods on regular grids as performed, e.g., in \cite{cahnhilliard}.
For boundary pixels of the image we use an analogue of Neumann zero boundary conditions, i.e., we assume the image is extended constantly.
This results in a total of $225 \cdot 400 = 90000$ hyperedges for the unoriented hypergraph, where each hyperedge $e_q \in \mathcal{E}_H$ can be directly associated with the corresponding image pixel.

\begin{figure}[htb]
\begin{subfigure}[t]{.48\textwidth}
\centering\includegraphics[width=\linewidth]{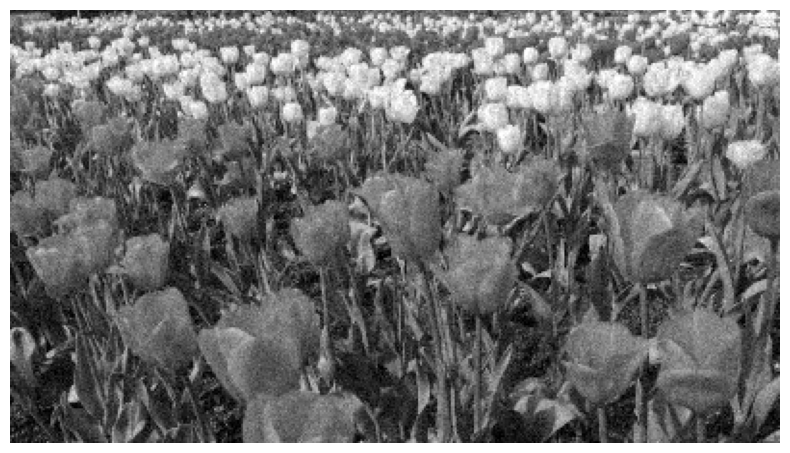}
\caption{$\tau=0.1, \mathbf{\lambda = 10}$, $100$ iterations}
\end{subfigure}
\begin{subfigure}[t]{.48\textwidth}
\centering\includegraphics[width=\linewidth]
{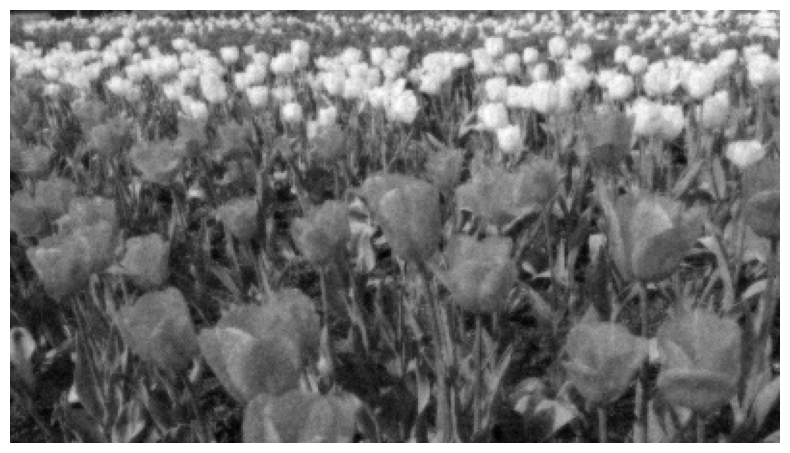}
\caption{$\mathbf{\tau=0.01}, \lambda = 0$, $100$ iterations}
\end{subfigure}\\
\begin{subfigure}[t]{.48\textwidth}
\includegraphics[width=\linewidth]{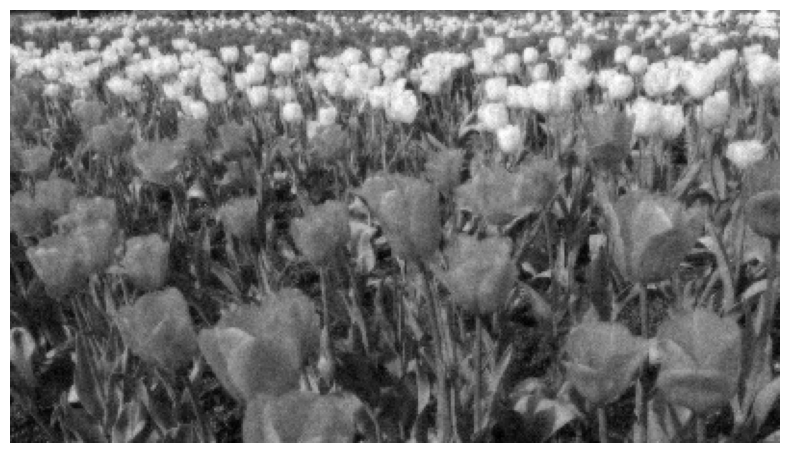}
\caption{$\tau=0.1, \mathbf{\lambda = 1}$, $100$ iterations}
\end{subfigure}
\begin{subfigure}[t]{.48\textwidth}
\centering\includegraphics[width=\linewidth]
{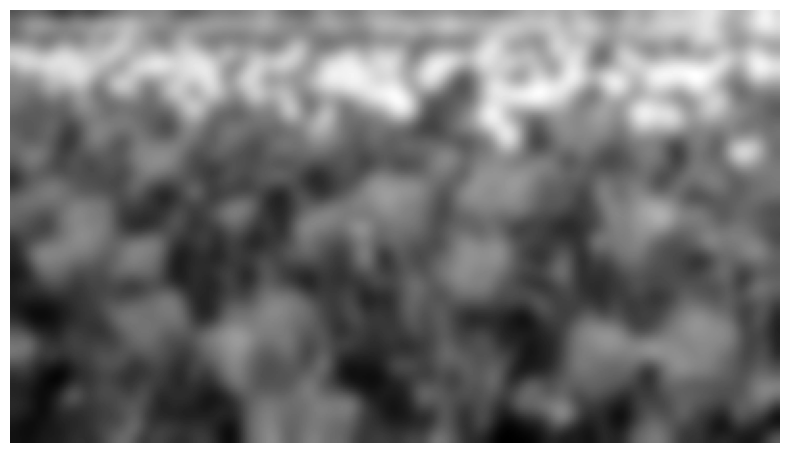}
\caption{$\mathbf{\tau=0.25}, \lambda = 0$, $100$ iterations}
\end{subfigure}\\
\begin{subfigure}[t]{.48\textwidth}
\centering\includegraphics[width=\linewidth]{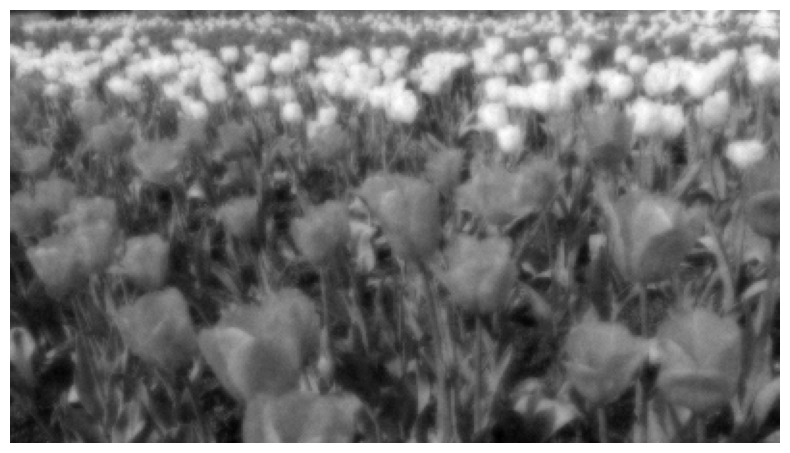}
\caption{$\tau=0.1, \mathbf{\lambda = 0.1}$, $100$ iterations}
\end{subfigure}
\begin{subfigure}[t]{.48\textwidth}
\centering\includegraphics[width=\linewidth]
{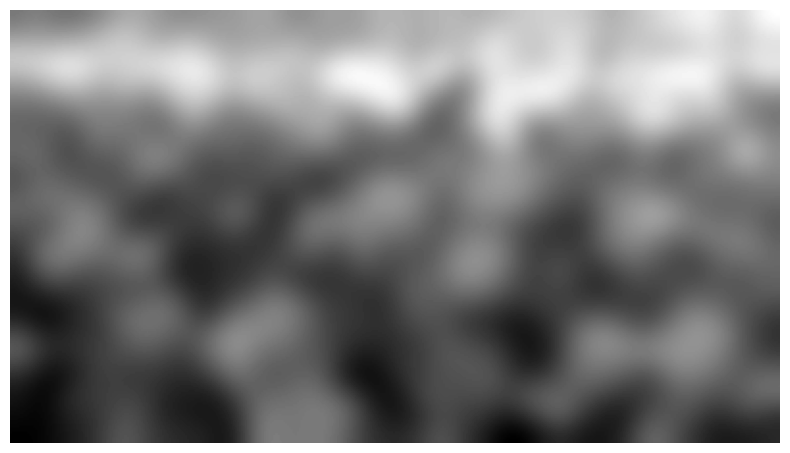}
\caption{$\mathbf{\tau=1}, \lambda = 0$, $100$ iterations}
\end{subfigure}\\
\begin{subfigure}[t]{.48\textwidth}
\centering\includegraphics[width=\linewidth]{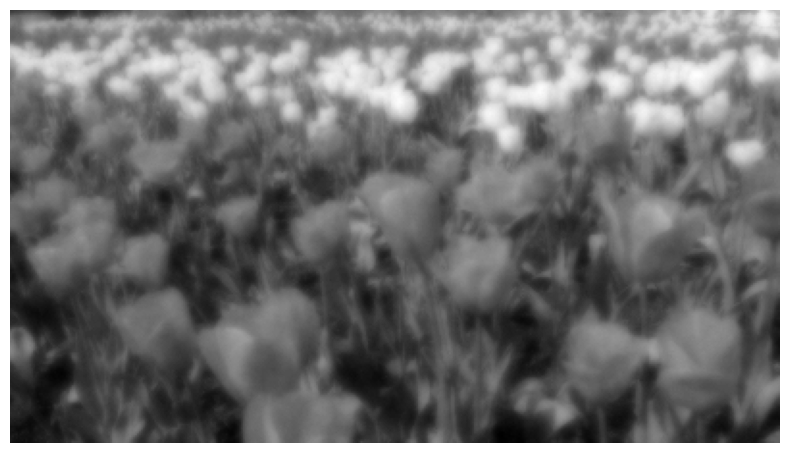}
\caption{$\tau=0.1, \mathbf{\lambda = 0.01}$, $100$ iterations}
\end{subfigure}
\begin{subfigure}[t]{.48\textwidth}
\centering\includegraphics[width=\linewidth]
{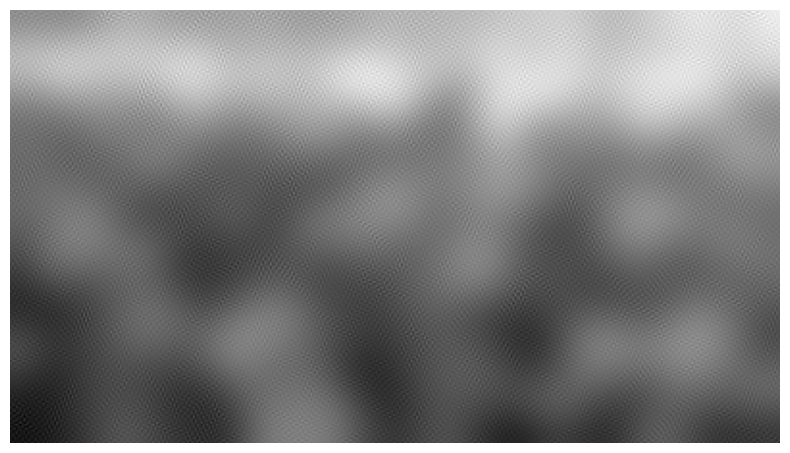}
\caption{$\mathbf{\tau=2}, \lambda = 0$, $100$ iterations}
\end{subfigure}\\
\caption{Solution of the initial value problem \eqref{eq:initial_value_problem_averaging} for the average hypergraph operator on a \textbf{local hypergraph} with data fidelity term (left) and without data fidelity term (right).}
\label{fig:local_image_processing}
\end{figure}
In the first case, we compare the results of using the iterative scheme \eqref{eq:iterative_scheme_averaging} with and without data fidelity term, for which we each performed $100$ iterations in our numerical experiments.
To investigate the influence of the data fidelity term, we first fixed the time step size as $\tau := 0.1$ and varied the regularization parameter $\lambda > 0$.
The left column of Figure \ref{fig:local_image_processing} shows that with decreasing value of $\lambda > 0$, the smoothing effect of the vertex averaging operator increases, leading to less noisy images.
On the other hand, the edges of image features get more and more blurry as can be expected in this case.
In another setting we remove the data fidelity term by setting $\lambda := 0$ and hence we investigate the corresponding evolution equation of the vertex averaging operator.
Here, we varied the time step size parameter $\tau > 0$ to compare different results for a fixed number of iterations.
As can be seen in the right column of Figure \ref{fig:local_image_processing}, we recover the scale spaces of the operator $\overline{\Delta_v}$.
With increasing time step size $\tau > 0$, we observe more and more coarse image features induced by the local averaging effect of the operator.

In the second numerical experiment we perform \textbf{nonlocal image processing} by constructing the hyperedges of the unoriented hypergraph not from the local neighborhood of an image pixel, but by regarding the pixel intensities of the image.
By this we are able to assemble vertices in hyperedges that correspond to image pixels which can be anywhere in the image and hence we gain a nonlocal vertex averaging operator.
In particular, we construct for each vertex $v_i \in \mathcal{V}$ a hyperedge containing all vertices for which the value of the vertex function $f$ is similar.
For this we choose a threshold $\epsilon > 0$ and add all vertices $v_j \in \mathcal{V}$ to the hyperedge induced by the vertex $v_i \in \mathcal{V}$ for which the distance is small enough, i.e.,
\begin{equation}
|f(v_i) - f(v_j)| < \epsilon.
\end{equation}
We chose $\epsilon := 40$ for the experiments presented in this setting.
With this approach we nonlocally group image pixels with similar grayscale intensities.
Since pixels with equal grayscale intensity lead to similar or even equal hyperedges, we treat hyperedges uniquely, i.e., without any duplicates in $\mathcal{H}(\mathcal{E})$.

\begin{figure}[htb]
\begin{subfigure}[t]{.48\textwidth}
\centering\includegraphics[width=\linewidth]{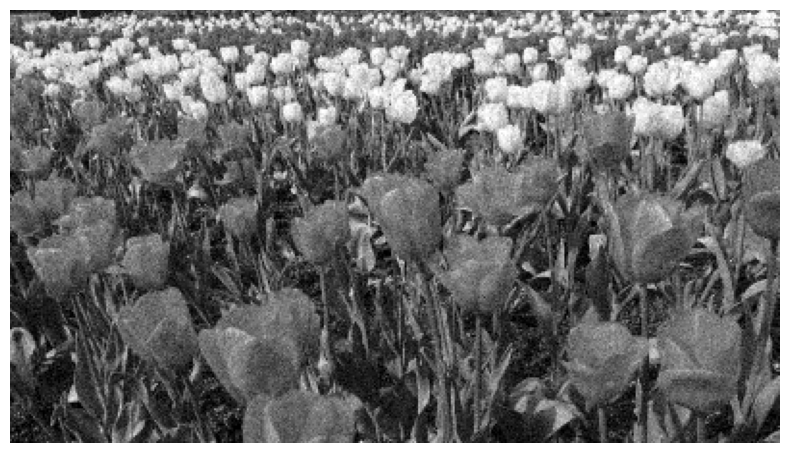}
\caption{$\tau=1, ~\mathbf{\lambda = 1}$, $100$ iterations}
\end{subfigure}
\begin{subfigure}[t]{.48\textwidth}
\centering\includegraphics[width=\linewidth]
{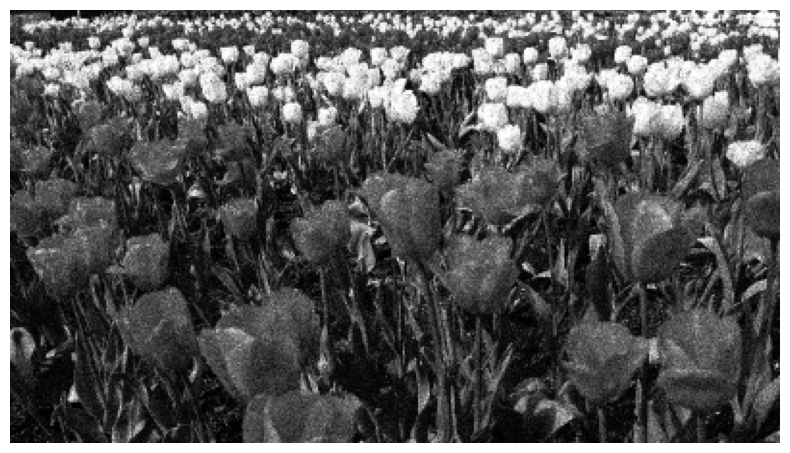}
\caption{$\mathbf{\tau=0.01}, ~\lambda = 0$, $100$ iterations}
\end{subfigure}\\
\begin{subfigure}[t]{.48\textwidth}
\centering\includegraphics[width=\linewidth]{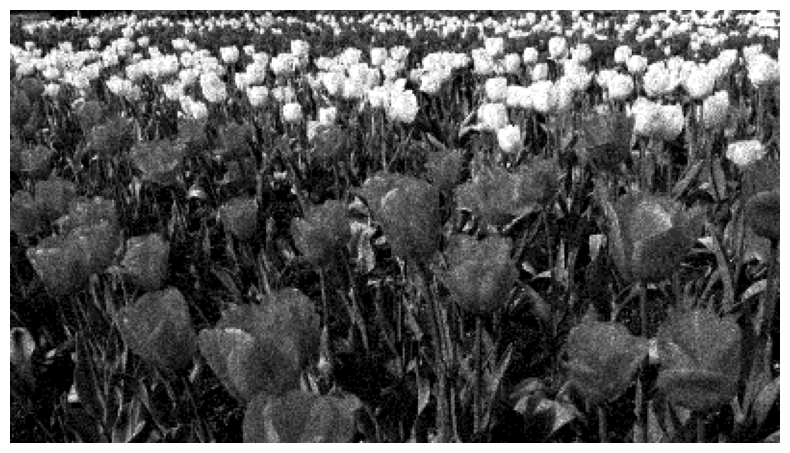}
\caption{$\tau=1, ~\mathbf{\lambda = 0.005}$, $100$ iterations}
\end{subfigure}
\begin{subfigure}[t]{.48\textwidth}
\centering\includegraphics[width=\linewidth]
{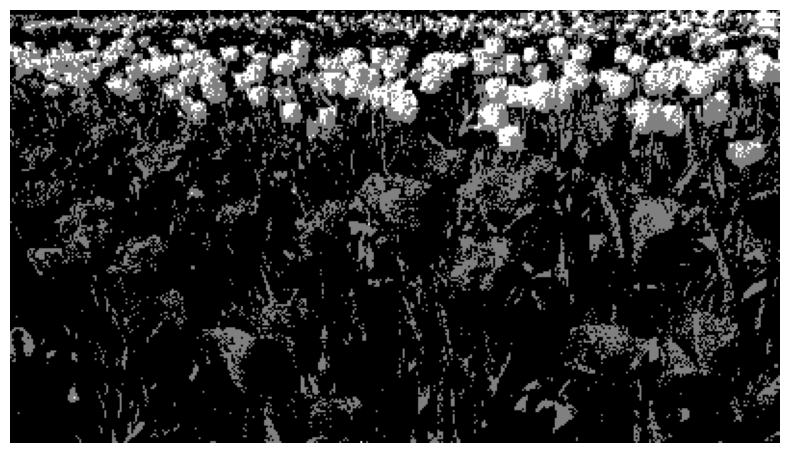}
\caption{$\mathbf{\tau=0.6}, ~\lambda = 0$, $100$ iterations}
\end{subfigure}\\
\begin{subfigure}[t]{.48\textwidth}
\centering\includegraphics[width=\linewidth]{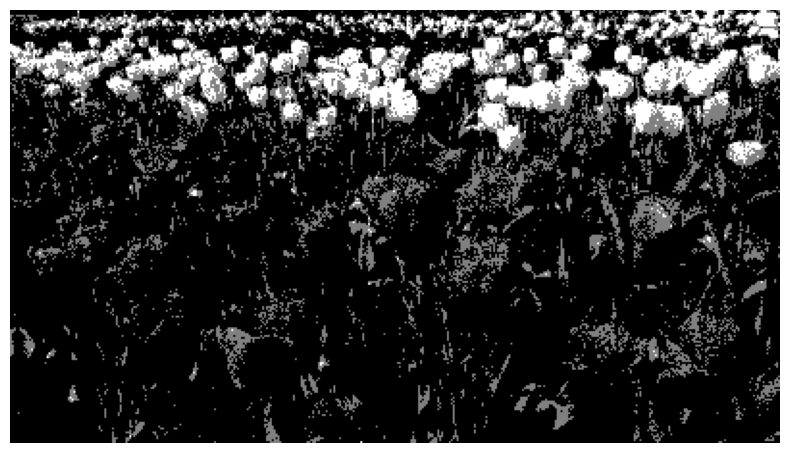}
\caption{$\tau=1, ~\mathbf{\lambda = 0.001}$, $100$ iterations}
\end{subfigure}
\begin{subfigure}[t]{.48\textwidth}
\centering\includegraphics[width=\linewidth]
{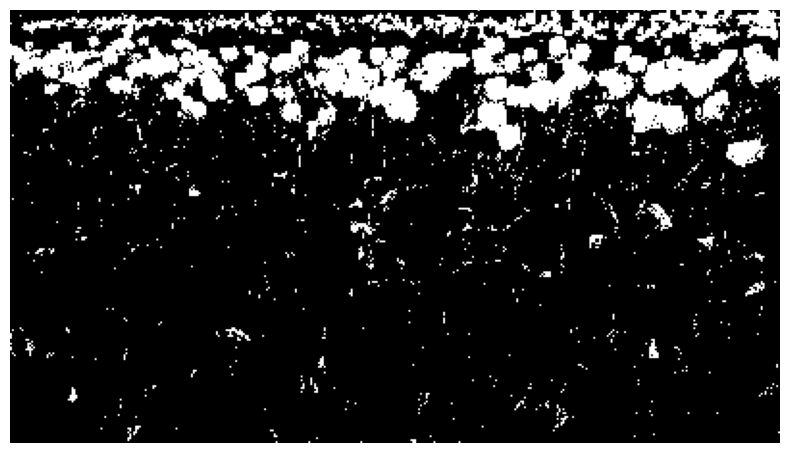}
\caption{$\mathbf{\tau=0.8}, ~\lambda = 0$, $100$ iterations}
\end{subfigure}\\
\begin{subfigure}[t]{.48\textwidth}
\centering\includegraphics[width=\linewidth]{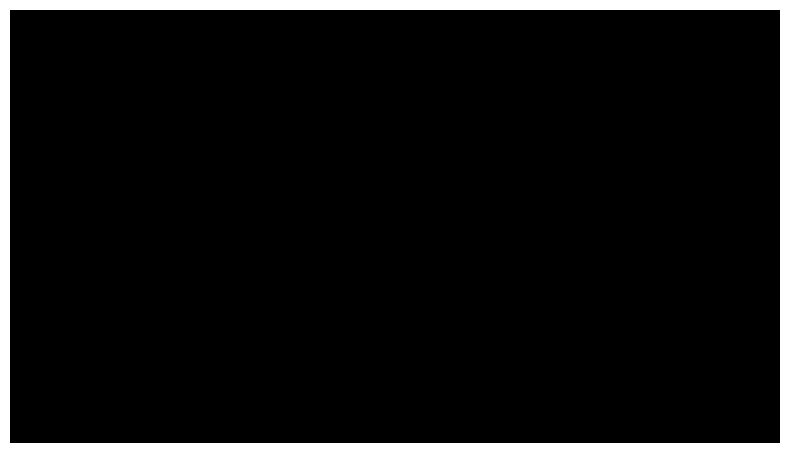}
\caption{$\tau=1, ~\mathbf{\lambda = 0.0001}$, $100$ iterations}
\end{subfigure}
\begin{subfigure}[t]{.48\textwidth}
\centering\includegraphics[width=\linewidth]
{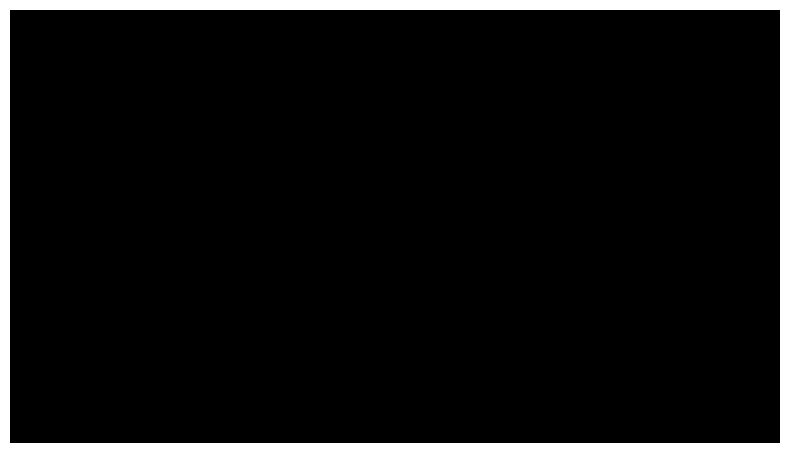}
\caption{$\mathbf{\tau=0.8}, ~\lambda = 0$, $100$ iterations}
\end{subfigure}
\caption{Solution of the initial value problem \eqref{eq:initial_value_problem_averaging} for the average hypergraph operator on a \textbf{nonlocal hypergraph} with data fidelity term (left) and without data fidelity term (right).}
\label{fig:nonlocal_image_processing}
\end{figure}
As in the case of local image processing described above, we compared the results of using the iterative scheme \eqref{eq:iterative_scheme_averaging} with and without data fidelity term, for which we performed $100$ iterations in our
numerical experiments.
We started our experiments by fixing the time step size as $\tau := 1$ and varying the regularization parameter $\lambda > 0$ in order to investigate the influence of the data fidelity term.
The left column in Figure \ref{fig:nonlocal_image_processing} shows that with decreasing value of $\lambda > 0$, the results deviate more and more from the initial image. Furthermore, one can observe that the amount of different pixel intensities decreases more and more until the resulting image shows only one grayscale intensity in the last row.
At the same time, edges of image features stay sharp as there is no averaging operation across the boundaries of image regions with strongly varying grayscale intensities.
Secondly, we again removed the influence of the data fidelity term entirely by setting $\lambda := 0$.
By varying the time step size parameter $\tau > 0$, we can compare different scales of the resulting scale space for a fixed number of iterations of the nonlocal hypergraph vertex averaging operator $\overline{\Delta_v}$.
As in the experiment with data fidelity, we observe that the amount of different pixel intensities decreases rapidly with increasing time step size $\tau > 0$, leading to a grouping of image pixels into image regions with similar grayscale intensities, until eventually all image pixels have the same grayscale value in the last row.
This expected behaviour can be leveraged for other image processing tasks in which grouping of image pixels is needed, e.g., in \textit{image compression} or \textit{segmentation}.
\section{Conclusion}
 
In this paper we derived {various variants of differential operators and} a family of $p$-Laplacian operators on hypergraphs, which generalize known graph and hypergraph operators from literature. In particular, we considered gradient, adjoint and $p$-Laplacian definitions both in the case of oriented as well as unoriented hypergraphs.

The resulting operators on oriented hypergraphs and the associated scale space flows can be employed for modelling information flows or performing spectral analysis in social networks, where we can directly incorporate group relationships via hyperarcs. Moreover, the proposed averaging operators and $p$-Laplacians for unoriented hypergraphs enable performing local and nonlocal image processing with results that can be used for segmentation tasks.
Preliminary results indicate a great potential for future research.  Interesting further questions, in addition to a more detailed study of spectral clustering, are e.g. the relation between hypergraph gradients and higher-order methods for partial differential equations or the definition of distance functions on hypergraphs via eigenfunctions of the infinity-Laplacian.

Another promising direction is to investigate the influence of non-constant weight functions of the hypergraphs used in our numerical experiments. In particular, we propose to define the weights of hyperedges based on the variance of the vertex function values for the vertices included in the respective hyperedge. This should further improve the results of both local and nonlocal image processing.

Due to the overarching success of learning-based methods in many areas of image processing and data analysis it seems obvious that further developments in this directions are in place for hypergraph structures. We believe that our work can provide a foundation for hypergraph neural networks generalizing the recently celebrated graph neural network structures with unforeseen opportunities.
%

\appendix

\section{Proof of Theorem \ref{thm:adjoint}}

\begin{proof}\ \\
   {Given an unoriented hypergraph $UH = \left(\mathcal{V}, \mathcal{E}_H\right)$ with vertex weight functions $w_I$ and $w_G$, and hyperedge weight functions $W_I$ and $W_G$, a vertex function $f \in \mathcal{H}\left(\mathcal{V}\right)$, and a hyperedge function $F \in \mathcal{H}\left(\mathcal{E}_H\right)$ then using the definitions of the inner product on $\mathcal{H}\left(\mathcal{E}_H\right)$ and of the vertex gradient $\nabla_v$ results in:\\}
    
    {\small$\begin{aligned}[t]
        {\langle G, \nabla_v f \rangle}_{\mathcal{H}\left(\mathcal{E}_H\right)} = & \sum_{e_q \in \mathcal{E}_H} W_I \left(e_q\right)^\beta G\left(e_q\right) \nabla_v f \left(e_q\right) &\\
    \end{aligned}$\\
    $\begin{aligned}[t]
        = & \sum_{e_q \in \mathcal{E}_H} W_I \left(e_q\right)^\beta G\left(e_q\right) W_G \left(e_q\right)^\gamma &\\
        & \qquad \left(\left(\sum_{v_i \in \mathcal{V}} \delta\left(v_i, e_q\right) w_I \left(v_i\right)^\alpha w_G \left(v_i\right)^\epsilon f\left(v_i\right)\right) - \left\lvert e_q\right\rvert w_I \left(v_{\tilde{q}}\right)^\alpha w_G \left(v_{\tilde{q}}\right)^\eta f\left(v_{\tilde{q}}\right)\right) &\\
        = & \sum_{e_q \in \mathcal{E}_H} W_I \left(e_q\right)^\beta G\left(e_q\right) W_G \left(e_q\right)^\gamma \left(\sum_{v_i \in \mathcal{V}} \delta\left(v_i, e_q\right) w_I \left(v_i\right)^\alpha w_G \left(v_i\right)^\epsilon f\left(v_i\right)\right) &\\
        & \qquad \quad - W_I \left(e_q\right)^\beta G\left(e_q\right) W_G \left(e_q\right)^\gamma \left\lvert e_q\right\rvert w_I \left(v_{\tilde{q}}\right)^\alpha w_G \left(v_{\tilde{q}}\right)^\eta f\left(v_{\tilde{q}}\right) &\\
        = & \sum_{e_q \in \mathcal{E}_H} \sum_{v_i \in \mathcal{V}} W_I \left(e_q\right)^\beta G\left(e_q\right) W_G \left(e_q\right)^\gamma \delta\left(v_i, e_q\right) w_I \left(v_i\right)^\alpha w_G \left(v_i\right)^\epsilon f\left(v_i\right) &\\
        - & \sum_{e_r \in \mathcal{E}_H} W_I \left(e_r\right)^\beta G\left(e_r\right) W_G \left(e_r\right)^\gamma \left\lvert e_r\right\rvert w_I \left(v_{\tilde{r}}\right)^\alpha w_G \left(v_{\tilde{r}}\right)^\eta f\left(v_{\tilde{r}}\right) &\\
        = & \sum_{e_q \in \mathcal{E}_H} \sum_{v_i \in \mathcal{V}} W_I \left(e_q\right)^\beta G\left(e_q\right) W_G \left(e_q\right)^\gamma \delta\left(v_i, e_q\right) w_I \left(v_i\right)^\alpha w_G \left(v_i\right)^\epsilon f\left(v_i\right) &\\
        - & \sum_{e_r \in \mathcal{E}_H} \sum_{v_j \in \mathcal{V}} \tilde{\delta}\left(v_j, e_r\right) W_I \left(e_r\right)^\beta G\left(e_r\right) W_G \left(e_r\right)^\gamma \left\lvert e_r\right\rvert w_I \left(v_{\tilde{r}}\right)^\alpha w_G \left(v_{\tilde{r}}\right)^\eta f\left(v_{\tilde{r}}\right) &\\
    \end{aligned}$}\\

    The last reformulation holds true, because for each hyperedge $e_r \in \mathcal{E}_H$ there exists exactly one special vertex $v_j \in \mathcal{V}$ with $v_j = v_{\tilde{q}}$ and thus the additional sum in combination with the characteristic function has no effect.\\

    {\small $\begin{aligned}[t]
        = & \sum_{e_q \in \mathcal{E}_H} \sum_{v_i \in \mathcal{V}} \left(W_I \left(e_q\right)^\beta G\left(e_q\right) W_G \left(e_q\right)^\gamma \delta\left(v_i, e_q\right) w_I \left(v_i\right)^\alpha w_G \left(v_i\right)^\epsilon f\left(v_i\right)\right. &\\
        & \qquad \qquad \quad - \left. \tilde{\delta}\left(v_i, e_q\right) W_I \left(e_q\right)^\beta G\left(e_q\right) W_G \left(e_q\right)^\gamma \left\lvert e_q\right\rvert w_I \left(v_{\tilde{q}}\right)^\alpha w_G \left(v_{\tilde{q}}\right)^\eta f\left(v_{\tilde{q}}\right)\right) &\\
        = & \sum_{v_i \in \mathcal{V}} \sum_{e_q \in \mathcal{E}_H} \left(W_I \left(e_q\right)^\beta G\left(e_q\right) W_G \left(e_q\right)^\gamma \delta\left(v_i, e_q\right) w_I \left(v_i\right)^\alpha w_G \left(v_i\right)^\epsilon f\left(v_i\right)\right. &\\
        & \qquad \qquad \quad - \left. \tilde{\delta}\left(v_i, e_q\right) W_I \left(e_q\right)^\beta G\left(e_q\right) W_G \left(e_q\right)^\gamma \left\lvert e_q\right\rvert w_I \left(v_{\tilde{q}}\right)^\alpha w_G \left(v_{\tilde{q}}\right)^\eta f\left(v_{\tilde{q}}\right)\right) &\\
        = & \sum_{v_i \in \mathcal{V}} \sum_{e_q \in \mathcal{E}_H} W_I \left(e_q\right)^\beta W_G \left(e_q\right)^\gamma G\left(e_q\right) &\\
        & \qquad \qquad \left(\delta\left(v_i, e_q\right) w_I \left(v_i\right)^\alpha w_G \left(v_i\right)^\epsilon f\left(v_i\right) - \tilde{\delta}\left(v_i, e_q\right) \left\lvert e_q\right\rvert w_I \left(v_{\tilde{q}}\right)^\alpha w_G \left(v_{\tilde{q}}\right)^\eta f\left(v_{\tilde{q}}\right)\right) &\\
        = & \sum_{v_i \in \mathcal{V}} \sum_{e_q \in \mathcal{E}_H} W_I \left(e_q\right)^\beta W_G \left(e_q\right)^\gamma G\left(e_q\right) &\\
        & \qquad \qquad \left(\delta\left(v_i, e_q\right) w_G \left(v_i\right)^\epsilon - \tilde{\delta}\left(v_i, e_q\right) \left\lvert e_q\right\rvert w_G \left(v_i\right)^\eta\right) w_I \left(v_i\right)^\alpha f\left(v_i\right) &\\
    \end{aligned}$}\\

    The last equality is based on the following argument: For a vertex $v_i \in \mathcal{V}$ and a hyperedge $e_q \in \mathcal{E}_H$ there are four cases:
    \begin{itemize}
        \item[1)] $\delta\left(v_i, e_q\right) = \tilde{\delta}\left(v_i, e_q\right) = 0 \quad \Longrightarrow$\\
        $\delta\left(v_i, e_q\right) w_I \left(v_i\right)^\alpha w_G \left(v_i\right)^\epsilon f\left(v_i\right) - \tilde{\delta}\left(v_i, e_q\right) \left\lvert e_q\right\rvert w_I \left(v_{\tilde{q}}\right)^\alpha w_G \left(v_{\tilde{q}}\right)^\eta f\left(v_{\tilde{q}}\right) = 0$\\
        \item[2)] $\delta\left(v_i, e_q\right) = \tilde{\delta}\left(v_i, e_q\right) = 1 \quad \Longrightarrow \quad v_i = v_{\tilde{q}} \quad \text{and}$\\
        $\delta\left(v_i, e_q\right) w_I \left(v_i\right)^\alpha w_G \left(v_i\right)^\epsilon f\left(v_i\right) - \tilde{\delta}\left(v_i, e_q\right) \left\lvert e_q\right\rvert w_I \left(v_{\tilde{q}}\right)^\alpha w_G \left(v_{\tilde{q}}\right)^\eta f\left(v_{\tilde{q}}\right) =$\\
        $w_I \left(v_i\right)^\alpha w_G \left(v_i\right)^\epsilon f\left(v_i\right) - \left\lvert e_q\right\rvert w_I \left(v_i\right)^\alpha w_G \left(v_i\right)^\eta f\left(v_i\right) =$\\
        $\left(w_G \left(v_i\right)^\epsilon - \left\lvert e_q\right\rvert w_G \left(v_i\right)^\eta\right) w_I \left(v_i\right)^\alpha f\left(v_i\right) =$\\
        \item[3)] $\delta\left(v_i, e_q\right) = 1 \quad \text{and} \quad \tilde{\delta}\left(v_i, e_q\right) = 0 \quad \Longrightarrow$\\
        $\delta\left(v_i, e_q\right) w_I \left(v_i\right)^\alpha w_G \left(v_i\right)^\epsilon f\left(v_i\right) - \tilde{\delta}\left(v_i, e_q\right) \left\lvert e_q\right\rvert w_I \left(v_{\tilde{q}}\right)^\alpha w_G \left(v_{\tilde{q}}\right)^\eta f\left(v_{\tilde{q}}\right) =$\\
        $\delta\left(v_i, e_q\right) w_I \left(v_i\right)^\alpha w_G \left(v_i\right)^\epsilon f\left(v_i\right)$\\
        \item[4)] $\delta\left(v_i, e_q\right) = 0 \quad \text{and} \quad \tilde{\delta}\left(v_i, e_q\right) = 1 \quad \text{impossible by definition}$\\
    \end{itemize}
    Hence the following equality holds true in all four cases, which verifies the feasibility of the reformulation above:\\
    $\delta\left(v_i, e_q\right) w_I \left(v_i\right)^\alpha w_G \left(v_i\right)^\epsilon f\left(v_i\right) - \tilde{\delta}\left(v_i, e_q\right) \left\lvert e_q\right\rvert w_I \left(v_{\tilde{q}}\right)^\alpha w_G \left(v_{\tilde{q}}\right)^\eta f\left(v_{\tilde{q}}\right) =$\\
    $\left(\delta\left(v_i, e_q\right) w_G \left(v_i\right)^\epsilon - \tilde{\delta}\left(v_i, e_q\right) \left\lvert e_q\right\rvert w_G \left(v_i\right)^\eta\right) w_I \left(v_i\right)^\alpha f\left(v_i\right)$\\

    With this equality we obtain the desired result:\\

    {\small $\begin{aligned}[t]
        & \sum_{v_i \in \mathcal{V}} \sum_{e_q \in \mathcal{E}_H} W_I \left(e_q\right)^\beta W_G \left(e_q\right)^\gamma G\left(e_q\right) &\\
        & \qquad \qquad \left(\delta\left(v_i, e_q\right) w_G \left(v_i\right)^\epsilon - \tilde{\delta}\left(v_i, e_q\right) \left\lvert e_q\right\rvert w_G \left(v_i\right)^\eta\right) &\\
        = & \sum_{v_i \in \mathcal{V}} w_I \left(v_i\right)^\alpha f\left(v_i\right) \sum_{e_q \in \mathcal{E}_H} \left(\delta\left(v_i, e_q\right) w_G \left(v_i\right)^\epsilon - \tilde{\delta}\left(v_i, e_q\right) \left\lvert e_q\right\rvert w_G \left(v_i\right)^\eta\right) &\\
        & \qquad \qquad \qquad \qquad \qquad \quad W_I \left(e_q\right)^\beta W_G \left(e_q\right)^\gamma G\left(e_q\right) &\\
        = & \sum_{v_i \in \mathcal{V}} w_I \left(v_i\right)^\alpha f\left(v_i\right) \nabla_v^* G\left(v_i\right) = {\langle f, \nabla^*_v G \rangle}_{\mathcal{H}\left(\mathcal{V}\right)} &\\
    \end{aligned}$}\\

    Therefore, the equality ${\langle G, \nabla_v f \rangle}_{\mathcal{H}\left(\mathcal{E}_H\right)} = {\langle f, \nabla^*_v G \rangle}_{\mathcal{H}\left(\mathcal{V}\right)}$ holds true for all vertex functions $f \in \mathcal{H}\left(\mathcal{V}\right)$ and all hyperedge functions $G \in \mathcal{H}\left(\mathcal{E}_H\right)$.
\end{proof}


\begin{thebibliography}{8}
    \bibitem{arnaboldi2017online}
    Arnaboldi, V., Conti, M., Passarella, A., Dunbar, R.: Online social networks and information diffusion: The role of ego networks. Online Social Networks and Media \textbf{1}, 44--55 (2017)

    \bibitem{bungertburger}
    Bungert, L.,  Burger, M.  Asymptotic profiles of nonlinear homogeneous evolution equations of gradient flow type. Journal of Evolution Equations \textbf{20}, 1061-1092 (2020).

    \bibitem{cahnhilliard}Burger, M., He, L., Sch\"onlieb, C. B. (2009). Cahn–Hilliard inpainting and a generalization for grayvalue images. SIAM Journal on Imaging Sciences, 2(4), 1129-1167
    
    \bibitem{chamley2013models}
    Chamley, C., Scaglione, A., Li, L.: Models for the diffusion of beliefs in social networks: An overview. IEEE Signal Processing Magazine \textbf{30}(3), 16--29 (2013)

    \bibitem{DiGiovanni}Di Giovanni, F., Rowbottom, J., Chamberlain, B. P., Markovich, T., Bronstein, M. M.: Graph neural networks as gradient flows. arXiv preprint arXiv:2206.10991 (2022)

    \bibitem{elmoataz2015p}
    Elmoataz, A., Toutain, M., Tenbrinck, D.: On the p-Laplacian and infinity-Laplacian on graphs with applications in image and data processing. SIAM Journal on Imaging Sciences \textbf{8}(4), 2412--2451 (2015)

    \bibitem{masterarbeit}
    Fazeny, A.: $p$-Laplacian Operators on Hypergraphs. Master thesis at FAU Erlangen-Nürnberg,  \url{https://gitlab.com/arianefazeny/hypergraph_p-laplace/-/raw/main/p-Laplacian_Hypergraphs_Fazeny_Ariane.pdf} (2023)


    \bibitem{fazeny2023}
    Fazeny, A., Tenbrinck, D., Burger, M.: Hypergraph p-Laplacians, Scale Spaces, and Information Flow in Networks. Proceedings on 9th International Conference on Scale Space and Variational Methods in Computer Vision, 677--690 (2023)


    \bibitem{jost2019hypergraph}
    Jost, J., Mulas, R.: Hypergraph Laplace operators for chemical reaction networks. Advances in mathematics 351, 870--896 (2019)

    \bibitem{jost2021plaplaceoperators}
    Jost, J., Mulas, R., Zhang, D.: p-Laplace Operators for Oriented Hypergraphs. Vietnam Journal of Mathematics Oct (2021)  

    \bibitem{snapnets}
    Leskovec J., Krevl A.: {SNAP Datasets}: {Stanford} Large Network Dataset Collection, \url{https://snap.stanford.edu/data/twitter-2010.html}. Last accessed 5 Oct 2022

    \bibitem{li2018submodular}
    Li, P., Milenkovic, O.: Submodular hypergraphs: p-laplacians, cheeger inequalities and spectral clustering. International Conference on Machine Learning, 3014--3023 (2018)

    \bibitem{majeed2020graph}
    Majeed, A., Rauf, I.: Graph theory: A comprehensive survey about graph theory applications in computer science and social networks. Inventions \textbf{5}(1), 10 (2020)

    \bibitem{mulas2022random}
    Mulas, R., Kuehn, C., B{\"o}hle, T., Jost, J.: Random walks and Laplacians on hypergraphs. Discrete Applied Mathematics 317, 26--41 (2022)

    \bibitem{consensus}
    Neuhäuser, L., Lambiotte, R., Schaub, M.: Consensus dynamics and opinion formation on hypergraphs. Springer International Publishing, 347--376 (2022)

    \bibitem{solomon2015} Solomon, J.  PDE approaches to graph analysis. arXiv preprint arXiv:1505.00185 (2015).

    \bibitem{stankovic2019} Stanković, L., Daković, M., Sejdić, E.: Introduction to graph signal processing. Vertex-Frequency Analysis of Graph Signals, 3--108 (2019)
    
    \bibitem{turcotte2015news}
    Turcotte, J., York, C., Irving, J., Scholl, R., Pingree, R.: News recommendations from social media opinion leaders: Effects on media trust and information seeking. Journal of computer-mediated communication \textbf{20}(5), 520--535 (2015)

    \bibitem{williamson2010lists}
    Williamson, S., Bender, E.: Lists, Decisions and Graphs. (2010)

    \bibitem{zanette}
    Zanette, D. H: Beyond networks: Opinion formation in triplet-based populations. Philosophical Transactions of the Royal Society A: Mathematical, Physical and Engineering Sciences, 367, 3311--3319  (2009)

    \bibitem{zhang2019signal}
    Zhang, S., Ding Z., Cui S.: Introducing hypergraph signal processing: Theoretical foundation and practical applications. IEEE Internet of Things Journal \textbf{7}.1, 639--660, (2019)

    \bibitem{zhou2006learning}
    Zhou, D., Huang, J., Sch{\"o}lkopf, B.: Learning with hypergraphs: Clustering, classification, and embedding. Advances in neural information processing systems \textbf{19} (2006)



    

\end{thebibliography}
\end{document}